# The life-cycle of star formation in distant clusters

A. J. Barger,[1] A. Aragón-Salamanca,[1] R. S. Ellis,[1] W. J. Couch,[2] I. Smail,[3] and R. M. Sharples[4]

[1] *Institute of Astronomy, Madingley Road, Cambridge CB3 0HA, UK*
[2] *School of Physics, University of New South Wales, Sydney 2052, Australia*
[3] *The Observatories of the Carnegie Institution of Washington, 813 Santa Barbara St., Pasadena, CA 91101-1292, USA*
[4] *Department of Physics, University of Durham, South Road, Durham DH1 3LE, UK*

**ABSTRACT**

We analyse the detailed distribution of star-forming and post-starburst members in three distant ($z = 0.31$) galaxy clusters in terms of evolutionary sequences that incorporate secondary bursts of star formation on pre-existing stellar populations. Using the number density of spectroscopically-confirmed members on the $EW(H\delta)$ versus $B - R$ plane from existing data, and for a larger $K'$–limited sample on the $U - I$ versus $I - K'$ plane from newly-acquired infrared images, we demonstrate that the proportion of cluster members undergoing secondary bursts of star formation during the last $\sim 2$ Gyr prior to the epoch of observation is probably as high as 30 per cent of the member galaxies. A key observation leading to this conclusion is the high proportion of $H\delta$–strong galaxies in all three clusters. The evolutionary modelling, whilst necessarily approximate, returns the correct proportions of galaxies in various stages of the star formation cycle both in terms of spectral and colour properties. HST images for the three clusters indicate a high proportion of the active members show signs of interaction, whereas the $H\delta$–strong galaxies appear mainly to be regular spheroidals. We examine results from recent merger simulations in the context of the populations in these clusters and confirm that the merging of individual galaxies, triggered perhaps by the hierarchical assembly of rich clusters at this epoch, is consistent with the star formation cycle identified in our data. The implications of such a high fraction of active objects in cluster cores is briefly discussed.

## 1   INTRODUCTION

Much recent work in observational cosmology has focused on the mounting evidence for significant evolution in a subset of rich cluster galaxies over the past $\sim 5$ Gyr. Butcher & Oemler (1978) were the first to present photometric evidence for a significantly higher fraction of blue galaxies in distant clusters than in present-day examples. This 'Butcher-Oemler' (BO) effect has since been confirmed in observational programmes employing broadband photometry (Butcher & Oemler 1984; Couch & Newell 1984), multiband optical and infrared imaging (Couch et al. 1983; Ellis et al. 1985; MacLaren, Ellis & Couch 1988; Aragón-Salamanca, Ellis & Sharples 1991, hereafter AES), spectroscopy (Dressler & Gunn 1982, 1983; Lavery & Henry 1986; Couch & Sharples 1987, hereafter CS), and HST imaging (Couch et al. 1994; Dressler et al. 1994). These observations indicate that the BO effect is a widespread starburst-related phenomenon occurring in the rich cluster environment at redshifts of $z \geq 0.2$.

The first substantial spectroscopic investigations of moderate redshift clusters were presented by Dressler & Gunn (1982, 1983). Following this, important progress was made by CS when they undertook a spectroscopic study with the multi-object fiber (FOCAP) system on the 3.9-m Anglo-Australian Telescope (AAT) to investigate the nature of the blue galaxy excess in 3 $z = 0.31$ clusters. Their high-quality 4 Å resolution spectra of $\sim 150$ objects within the fields of the rich clusters AC103, AC114, and AC118 enabled them to assess the star-forming activity of the individual galaxy cluster members. They constructed a useful diagnostic diagram based on the distribution of rest-frame $H\delta$ $\lambda4103$ Å equivalent widths (hereafter $H\delta$) and reddening-corrected $B_J$-$R_F$ colours for the cluster members, which they then compared to predictions for model galaxies that evolve according to the Bruzual galaxy evolution code (Bruzual 1981, 1983). Their study revealed that large numbers of blue and red galaxies were or had recently been in unusual phases of star-forming activity, as indicated by the location of the galaxies on this diagram.

CS defined five spectral categories for their cluster galaxies according to location on the $H\delta$−colour plane. Blue galaxies lying along the nearby spiral galaxy sequence of increasing $H\delta$ strength with bluer colour were assumed to be normal spiral members. However, blue galaxies with emission-filled $H\delta$ lines were defined to be 'starburst' (SB) galaxies, and those with moderate to strong $H\delta$ absorption ($H\delta > 4$ Å) 'post-starburst' (PSG) galaxies. The SB types were interpreted as galaxies undergoing a short secondary burst of star formation at the time of observation, whereas the PSG types had recently completed such a phase. The red galaxy population appeared to be composed of galaxies with spectra and rest-frame colours equivalent to nearby E/S0's together with examples showing enhanced $H\delta$ absorption ($H\delta > 2$ Å). The latter were classed as '$H\delta$-strong' (HDS) galaxies and were interpreted as galaxies viewed $< 2$ Gyr after the truncation of either ongoing or bursting star formation.

Some confusion has arisen in the literature about the 'E+A' classification of galaxies, partly as a result of the wide range of signal/noise achieved in the various datasets. The 'E+A' terminology, as originally defined by Dressler & Gunn, refers to a spectral class of objects defined using Balmer absorption line strengths without regard to broadband colour; thus, broadly speaking, the 'E+A' class includes both PSG and HDS sources. In this paper we will avoid using the 'E+A' notation.

Couch et al. (1994) discuss the morphological nature of galaxies in the various spectral classes defined above from the first HST WFPC-1 images of AC114 and Abell 370. Although their sample is restricted by the small WFPC-1 field, they find that a high proportion of the SB and PSG galaxies show signs of interaction and/or disk structures, whereas the HDS galaxies are almost exclusively isolated spheroidals. Post-repair HST data are now being secured for many more clusters so that significant progress can be expected in understanding the morphological distributions of the various spectral classes.

Several workers (CS; MacLaren et al. 1988; AES) have proposed that the spectral and photometrically-defined galaxy classes might represent different stages within one cycle of secondary star formation. If correct, this might imply one underlying physical cause for the recent star formation in moderate redshift clusters and hence have far-reaching consequences. Such a process would have to be widespread in the centers of rich clusters at $z \geq 0.2$ but be noticeably absent in these regions today.

We can use stellar evolutionary codes to model the *frequency* of galaxies observed in the various spectral categories to test whether a single cycle of star formation can reproduce the CS spectroscopic cluster observations. If so, further quantitative constraints on the parameters governing the starbursts may result. The CS sample provides important constraints on our analysis; however, inasmuch as it is an $R$−band limited sample, it may be significantly biased by starburst activities. This bias can only be overcome with an infrared-selected sample in which luminosities are more indicative of long-lived stellar populations. To this end, we have obtained new deep infrared images of the 3 well-studied CS clusters and constructed $K'$−selected optical and infrared photometric catalogues for further testing of our model predictions in the colour–colour plane.

Our purpose in this paper is to analyse the numbers of galaxies in the various spectral and photometric categories for the 3 CS clusters. We use the isochrone spectral synthesis code of Bruzual & Charlot (1993, hereafter BC) to model the evolution of normal galaxies experiencing a secondary burst of star formation either before or during the time of observation. These models are used to generate evolutionary tracks and population densities for various parameter spaces. The predictions are compared with the available spectroscopic and photometric data. This is a particularly timely study given the morphological details available from HST, the $K'$−limited photometric data samples for reducing the starburst selection effects, and recent efforts by Mihos (1995) to simulate the *dynamical* evolution of merging galaxies in clusters.

The organization of the paper is as follows. In §2 we introduce the basic model assumptions. In §3 we illustrate the importance of allowing for the luminosity bias introduced during the active phase of a burst, and we discuss how deep infrared photometry may overcome selection biases affecting earlier optically-selected samples. We then develop methods for generating number density distributions for individual clusters by assuming that various proportions of spiral and early-type galaxies are observed stochastically during a continuous star-formation cycle. In §4 we compare the predictions with the CS spectroscopic cluster data. We then present the new deep infrared images of the 3 CS clusters and the reduction procedures used for building $K'$−selected optical and infrared photometric catalogues. We determine the $K'$−band cluster luminosity function and we present colour-magnitude and colour-colour diagrams. The $U-I$ versus $I-K'$ colour plane provides a further check on the burst cycle hypothesis. In §5 we examine the predictions of recent dynamical simulations in the context of HST morphologies for our cluster galaxies. This acts as an independent test of the secondary starburst cycle picture in the specific case where the star-formation activity is merger-induced. Our main results, their limitations, and possible alternatives are summarised in §6.

## 2    MODEL ASSUMPTIONS

We aim to test the hypothesis that galaxies undergoing secondary bursts of star formation can account for the various manifestations of recent star-forming activity seen in distant clusters. We test this by predicting the numbers of galaxies expected in the various spectral categories using model galaxy spectra generated according to the BC galaxy evolution code. The stellar evolutionary tracks in this code cover all relevant phases of evolution from the main sequence to the remnant phase. One possible drawback is that the models assume solar metallicity, thereby neglecting chemical evolution effects; however, as we are primarily concerned with testing short-term changes to old well-established galaxies, this should not be a major deficiency.

The BC models assume a given initial stellar mass function (IMF) and star-formation history and output as a func-tion of time observables such as spectral energy distributions (SED), absolute magnitudes, and colours. We adopt the Scalo IMF with lower and upper mass cutoffs of 0.1 and 125 $M_\odot$, respectively, for our models. We experimented with modelling nebular emission lines and continuum emission using recombination theory, but as the results did not vary significantly from those obtained using a continuum-subtracted spectrum of the nearby Sbc spiral NGC2997 (as in CS), we decided to adopt the latter. The NGC2997 spectrum is rescaled by the ratio of the predicted $H\beta$ Case B recombination flux to the NGC2997 $H\beta$ flux and added to the model absorption spectrum.

By experimenting with a variety of star-formation histories for their model galaxies, CS found that short-term 'burst' models were the most successful in reproducing the observed colours and $H\delta$ absorption features of their data. The abrupt change to a higher star formation rate (SFR) was an essential ingredient in generating model spectra similar to those in the SB category, while the observed PSG spectra could be reproduced by burst models viewed shortly after star formation had ceased. Although CS discussed the distribution of galaxies on their $H\delta-$colour plane, no account was taken of the relative numbers in the various regions.

Clearly there is ample scope for generating elaborate evolutionary models with numerous input parameters, few of which can be physically constrained. Our aim instead is to consider the simplest physical situation which is compatible with the data at hand. We begin by examining two basic forms of bursting model galaxies. These model galaxies are assumed to be viewed by an observer to whom the galaxies would have a redshift of $z = 0.31$, assuming galaxy formation at $z \sim 5$ and $H_0 = 50$ km s$^{-1}$ Mpc$^{-1}$ in an $\Omega = 1$ cosmology. In the 'old population+burst' (OPB) history, a brief second burst of star formation is added to an old population. BC's c-model, characterised by a single burst of star formation in the first billion years, is chosen for this class since it fits the data for local and moderate redshift early-type galaxies in the quiescent phase (AES). In the 'spiral population+burst' (SPB) history, a second burst is added to a galaxy which has undergone near-constant star formation during its lifetime. Bruzual (1981) introduced $\mu-$models to model such systems. These share an exponentially-decaying SFR with $\mu$ as the decay parameter, defined as the fraction of gas converted to stars in the first Gyr. Systems undergoing a constant SFR can be approximated by models with $\mu = 0.01$.

The basic variables in these two simplified cases are the same: (i) the *strength of the second burst* relative to the total integrated star formation and (ii) the *burst duration* in Gyr. We experimented with burst strengths equivalent to the conversion of $10 - 40$ per cent of the final stellar mass into stars during periods which varied from $0.1 - 1$ Gyr. In both the OPB and SPB cases all star formation is assumed to cease after the completion of the second burst.

Figures 1 contrast the OPB and SPB evolutionary tracks of the rest-frame $H\delta$ and the $B-R$ colour[*], as viewed by an observer to whom these galaxies would have a redshift of 0.31. In both cases $H\delta$ remains an important diagnostic of the burst for a considerable time after the burst. In the OPB case, before the second burst occurs the underlying model galaxy colours are reddening progressively, but the galaxy is transformed by the burst and does not return to its pre-burst state until several Gyr later. In the SPB case, the immediate effect of the burst, whilst significant, is less dramatic. As we assume the star formation is truncated after the burst, the post-burst evolution is relatively similar in both cases, i.e. the only opportunity to constrain the original spectral type is during the active period.

## 3 THE STAR-FORMATION CYCLE

We now wish to use the trends illustrated in Figs. 1 to make numerical predictions of how many galaxies are viewed in each of the SB:PSG:HDS and normal stages classed by CS. For such a comparison we must consider that a magnitude-limited sample is more sensitive to bursting galaxies than quiescent ones. When galaxies otherwise fainter than the selection limit undergo a strong burst of star formation they become more luminous, and some will be boosted into the sample. This bias is stronger at short wavelengths where the rest-frame light is dominated by the contribution from the young stellar populations. Knowledge of the quiescent luminosity function of the cluster is needed to properly calculate the number distributions for the $R-$limited case. In the $K-$band, however, the luminosities are more indicative of long-lived stellar populations; thus, luminosity boosting is much smaller. Figure 2 shows the observed $R-$ and $K'-$band absolute magnitudes, as viewed at a redshift of 0.31, for the model galaxies in Figs. 1. In §4 we will use our newly constructed $K'-$limited optical and infrared photometric samples to further test the model predictions obtained for the $R-$limited spectroscopic data.

The LF could be derived directly from the cluster photometry obtained by CS. However, given that such photometric data are affected by the very processes we wish to study, we prefer to determine a 'non-bursting' LF from alternative sources. The precise form of the LF is not critical to our subsequent analysis; in practice we make use of the Loveday et al. (1992) $b_J$ field LF. Since the $R$ passband closely measures the rest-frame $B-$band flux of a $z = 0.31$ galaxy, we can follow the procedure of AES and convert the local $b_J$ LF to an $R_{z=0.31}$ LF using the transformation $(M_{b_J})_{z=0} = (M_R)_{z=0.31} + 0.5905(M_B - M_V)_{z=0} + 0.5828$. A further important consideration here is that the LF be sampled fainter than the limit to which the spectral data are to be used.

In the simplest case we wish to examine, we assume that some fraction of the galaxies in a cluster evolve stochastically according to a template $H\delta$ versus time and colour versus time 'life-cycle' given by one of the curves in Figs. 1. For each model galaxy, Monte Carlo methods are used to assign a luminosity (and hence apparent magnitude) drawn from the adopted LF, a model template (and hence morphology), and the time of the second burst, constrained to occur no more than 1.8 Gyr prior to the epoch of observation. Objects undergoing bursts earlier than this will not show any effects of the burst by the epoch of observation. The last ingredi-ent is randomly distributed and used to apply a time shift to the template so that a continuous population of sources feeds the cycle. The observer then, in effect, observes galaxies at various stages along a single duty cycle. The luminosity brightening ($\Delta R$) associated with any specified model can be incorporated readily (c.f. Fig. 2) by simulating the process fainter than the actual magnitude limit and then applying the magnitude limit retrospectively.

[*] Whereas CS used photographic-based $B_J - R_F$ colours in their paper, we will adopt standard $B - R$ hereafter in our analysis and modelling. The difference is $< 0.02$ mag over the colour range of interest (Couch & Newell 1990)

**Figure 2.** The luminosity brightening in the observed $R-$ and $K'-$bands for an early-type galaxy undergoing a secondary burst of star formation, as viewed by an observer watching a cluster at $z = 0.31$. Note the smaller brightening in $K'$ suggests that $K'-$limited samples will better represent pre-burst populations.

Figures 3 show $H\delta - (B-R)$ model tracks for a selection of burst strengths and durations for (a) OPB galaxies and (b) SPB galaxies. The tracks are marked with time intervals measured from the beginning of the second burst to the time of observation. It can be seen that the *strength* of the burst governs both the extent of the excursion into the blue section of the plane and, to a lesser degree, the maximum depth of the Balmer absorption line in the post-burst phase. For a given burst strength, however, increasing the *duration* of the burst actually works in the opposite sense. This is because it is the contrast of the old and young populations which determines the excursions of galaxies on the $H\delta - (B-R)$ diagram. A short intense burst produces very blue galaxies. Beyond a few hundred Myr, the post-burst behaviour of all models is similar. A short burst produces identical activities regardless of the underlying type.

As an illustration of the technique, in Fig. 4 we show the distribution of expected sources constructed using the adopted cluster LF and assuming the population is composed (a) entirely of spheroidals, all of which are bursting according to the above OPB history or (b) entirely of spirals, all of which are bursting according to the SPB history. We have simulated galaxies intrinsically fainter than the CS limit of $R = 20$ and limited the data retrospectively once the

**Figure 1.** The evolution as a function of time of (a) the rest-frame equivalent widths of the Balmer absorption line $H\delta$, (b) the $B-R$ colours, and (c) the $U-I$ colours, as viewed by an observer watching a $z = 0.31$ cluster, for galaxies undergoing secondary bursts of star formation. The OPB notation refers to an early-type galaxy modelled by an initial single burst and supplemented by a secondary 10 per cent burst, and the SPB notation refers to a spiral galaxy modelled by a continuous star formation rate and supplemented by a secondary 10 per cent burst. In both cases all star formation ceases after the completion of the secondary burst.

**Figure 3.** The evolutionary tracks of the (a) OPB and (b) SPB star formation histories on the $EW(H\delta)$ versus $B - R$ plane, as viewed by an observer watching a $z = 0.31$ cluster. Time marks refer to intervals in Gyr from the time of the beginning of the second burst to the time of observation. The effects of varying the burst duration and strength percentages are shown.

**Figure 4.** Simulated distribution of galaxies in the $EW(H\delta)$ versus $B - R$ plane for a $z = 0.31$ cluster populated entirely by (a) spheroidals undergoing bursts via the OPB star formation history and (b) spirals undergoing bursts via the SPB star formation history. The simulated data include the effects of likely observational errors.

$\Delta R$ brightening has been included. For future comparisons with the data we incorporate observational errors assuming Gaussian distributions with $\sigma_{B-R} = 0.1$ and $\sigma_{H\delta} = 1.0$ Å, typical of the values in CS.

To facilitate comparisons with the observations, we divide the plane into regions, following CS. The plane is first split into two colour halves ('blue' and 'red') with the boundary at $B - R = 2.0$. Red galaxies with $H\delta \leq 3.0$ Å are defined as 'normal spheroidals', while those with $H\delta > 3.0$ Å are classed as 'HDS galaxies'. Among the blue galaxies, those with $H\delta < 2.0$ Å are considered 'SB galaxies', those with $2.0$ Å $\leq H\delta \leq 6.0$ Å are defined for convenience to be 'continuous star-forming galaxies' and can be associated with normal spirals. The remaining galaxies, with $H\delta > 6.0$ Å, are termed 'PSG'.

The precise boundaries of these regions are somewhat arbitrary but, provided the number densities in each region are consistently compared between the model predictions and data, our conclusions will be robust. Clearly if we are cautious in our definition of normal spirals, this might overestimate the proportion of unusual objects. It is important to emphasise here that we are primarily defining normal objects to be those with conventional star-formation histories. Objects outside this box may be found locally, but it is their rarity which makes them of value and our numerical comparisons will emphasise this directly.

In practice, it is likely that only a fraction of the early-type and spiral galaxies are involved in the life-cycle and so, for complete generality, we must finally introduce two more parameters: the relative starting proportions of early-types and spirals, and the fraction of each type undergoing the activity. At first sight this seems an unfortunate increase in the number of parameters, but in fact both are constrained by the observed numbers and colours of galaxies occupying the 'quiescent' regions of the $H\delta-$colour plane, which our models must also reproduce. Determining the fraction undergoing recent star-forming activity is a major goal of this work, since it may constrain the possible mechanisms.

## 4    DATA AND MODEL COMPARISONS

### 4.1    Comparisons with the $H\delta-$colour data

The observed datasets for each of the 3 clusters taken individually are too sparse for reliable comparisons. Accordingly, we have combined the data from all 3 CS clusters, noting that they represent a relatively homogeneous sample with identical redshifts, magnitude limits, and field areas, and with fairly similar blue fractions and optical richnesses. The principal differences lie in the cluster morphologies. AC114 is a cD-dominated system which may be relatively evolved, whereas AC118 and AC103 still show signs of substructure.

Figure 5a shows the $H\delta-$colour distribution for the combined $R < 20$ cluster data sample characterised by the 5 categories introduced in the previous section. The sample sizes are indicated by the numbers $N_{E/S0}$, $N_{SB}$, $N_{Spiral}$, $N_{PSG}$, and $N_{HDS}$. Given the small numbers involved, there is no strong difference in the trends seen from one cluster to another as indicated by the different symbols.

The aim now is to compare the relative numbers with the predictions of the stochastic 'continuous cycle' models described in §3. As can be seen from Figs. 3, the relative numbers of SB, PSG, and HDS objects are particularly sensitive indicators. Short bursts have a longer PSG phase than long bursts and are the only models capable of penetrating the blue end of the PSG box. Burst parameters of 0.1 Gyr and $10-20$ per cent burst strength were found to produce a good description of the colours and spectral properties of the starburst galaxies in the 3 clusters. We will adopt these parameters in what follows, recognising that our overall conclusions are not too sensitive to small changes in these parameters. All bursting galaxies enter the SB and HDS categories, but they remain in the HDS category much longer; thus, the total fraction of HDS sources is a good indicator of the overall proportion of bursting galaxies. So many HDS galaxies are observed in the 3 clusters that, taking into account the lifetime of this phase, it is clear that the bursting phenomenon is very widespread.

Figure 5b shows a model where 30 per cent of the members of a cluster composed of 60 per cent spheroidals and 40 per cent spirals are undergoing the star-formation cycle, following either the OPB or the SPB track. The $N_{HDS}$ frequency offers a way of determining the total fraction of burst objects to within $\pm 7$ per cent. In the figure we distinguish between the bursting and the non-bursting objects. The non-bursting object colours were obtained from folding local elliptical and spiral galaxy SED's with the relevant filters, assuming the galaxies were actually located at a redshift of 0.31. Since no information exists on the spiral population distribution in moderate redshift clusters, we based the non-bursting spiral distribution in the finer morphological classes on that found in the Coma cluster (Dressler 1980). We stress that the aim was not to constrain the non-bursting population distribution in detail but simply to show that the observations could be made consistent with sensible choices.

The sensitivity to the composition and bursting parameters is illustrated in Table 1 where two spheroidal:spiral percentage compositions (60:40 and 70:30) are considered. A range of bursting population percentages ($10-50$ per cent) are examined in the former case. Column 1 lists the cluster composition, column 2 the percentage of members undergoing secondary bursts, and columns 3-7 the means and sigmas for each of the various categories defined above. The data to compare with is listed in the top row of the table. Bursting populations in the range $20-30$ per cent give the best fits to all the numbers in the various spectral categories. $N_{HDS}$ places an upper bound on the bursting population to less than 40 per cent.

To simplify the analysis we have admittedly explored only a restricted set of reasonable star formation histories. In particular, we have assumed star formation is truncated after the secondary burst, even in disk galaxies. One might ask whether our conclusions would be affected if disk galaxies underwent starbursts and then returned to more normal patterns of star formation. Such a star-formation history might occur in cluster spirals which undergo perturbations milder than an actual merger. This seems unlikely given the absence of [O II] emission in the PSG population, a primary factor for considering truncated models. However, we find the resulting trajectory in the $H\delta$ versus $B-R$ plane for the case of continuing star formation is very similar to the SPB model considered previously, except that the maximum equivalent width is about 1 Å lower (from emission line infill) and the trajectory in Fig. 3b would terminate at $B-R < 2$. Thus, the contribution of galaxies with continuing star formation cannot be large because it would greatly underpopulate the HDS region. Additionally, objects that lie well away from our normal spiral box, in particular the PSG's at $H\delta > 8$Å, can only be explained by invoking quite massive secondary bursts of star formation ($\sim 20$ per cent for spirals).

In summary, therefore, the density of cluster members in various portions of the $H\delta-$colour plane for AC103, AC114, and AC118 is best reconciled with a secondary starburst cycle in which $\simeq 30$ per cent of the cluster population have undergone activity of this type within the last $\sim 2$ Gyr prior to the epoch of observation. Undoubtedly, the model chosen in Fig. 5b is not unique; nevertheless, our model simulations based on a variety of parameter choices lead us to the conclusion that a significant percentage of the cluster population must be involved.

**Table 1.** Model results.

| E/S0:Sp | % | $N_{E/S0}$ | $N_{Sp}$ | $N_{SB}$ | $N_{PSG}$ | $N_{HDS}$ | $N_{quies.}$ | $N_{blue}$ | $N_{UV}$ | $< \chi^2 >$ |
|---------|---|-----------|----------|----------|-----------|-----------|--------------|------------|----------|--------------|
|         |   | 59        | 17       | 10       | 6         | 20        | 107          | 71         | 24       |              |
| 60:40   | 10 | $81.6 \pm 4.9$ | $10.6 \pm 2.3$ | $1.2 \pm 1.3$ | $3.6 \pm 1.4$ | $15.0 \pm 3.8$ | $114.8 \pm 6.3$ | $63.1 \pm 6.0$ | $24.2 \pm 4.0$ | 25.60 |
| 60:40   | 20 | $74.8 \pm 4.1$ | $9.6 \pm 3.6$ | $2.8 \pm 1.6$ | $5.3 \pm 2.0$ | $19.6 \pm 3.6$ | $112.4 \pm 6.7$ | $63.1 \pm 6.6$ | $27.1 \pm 4.3$ | 18.40 |
| 60:40   | 30 | $66.6 \pm 5.1$ | $9.0 \pm 3.2$ | $4.8 \pm 2.2$ | $6.9 \pm 2.4$ | $24.8 \pm 4.6$ | $109.0 \pm 5.8$ | $62.2 \pm 4.9$ | $30.8 \pm 4.6$ | 16.48 |
| 60:40   | 40 | $58.0 \pm 5.3$ | $9.4 \pm 2.2$ | $5.2 \pm 2.4$ | $7.8 \pm 2.1$ | $31.5 \pm 4.0$ | $103.0 \pm 8.5$ | $67.8 \pm 6.0$ | $32.2 \pm 5.1$ | 20.68 |
| 60:40   | 50 | $52.9 \pm 3.7$ | $9.8 \pm 2.4$ | $5.7 \pm 2.1$ | $10.3 \pm 3.3$ | $33.3 \pm 2.4$ | $96.2 \pm 4.7$ | $70.2 \pm 5.1$ | $35.7 \pm 5.2$ | 28.83 |
| 70:30   | 30 | $67.5 \pm 2.8$ | $8.9 \pm 2.5$ | $4.2 \pm 2.2$ | $6.1 \pm 2.1$ | $25.3 \pm 2.5$ | $122.0 \pm 6.1$ | $55.1 \pm 6.2$ | $24.9 \pm 5.4$ | 19.38 |

## 4.2 New photometric data

As discussed earlier, optically-selected samples are inherently biased towards starbursting systems. Although we have attempted to correct for this bias in our preceding analyses, a complete $K$–limited spectroscopic sample with accurate $H\delta$ indices would strongly reduce the corrections necessary and hence yield more reliable results. In preparation for such a spectroscopic sample, we have obtained deep $K'$–band images with the IRIS infrared imager on the AAT (Figs. 6). We have used these data, together with $U-$ and $I$–band CCD images taken on the CTIO 4-m and ESO 3.5-m NTT, to compare long baseline colour-colour diagrams for the cluster data to further test our model predictions.

### 4.2.1 Infrared data

The near infrared data were obtained at the 3.9-m Anglo-Australian Telescope with the IRIS infrared camera using a $128 \times 128$ HgCdTe detector array manufactured by the Rockwell International Science Center. For a detailed description of the camera and observing procedures see Allen (1992) and references therein. The f/36 focal ratio was used, yielding a pixel scale of $0.79''$ pixel$^{-1}$. The observations were made through an intermediate band filter, called $K'$ for convenience, introduced to avoid excessive thermal background at longer wavelengths in the $K$ passband. Table 2 contains a log of the infrared observations. All nights were clear and photometric with adequate seeing ($1.1'' - 1.7''$).

In order to cover the area imaged with HST, we completed a $2 \times 2$ mosaic for each cluster with a $10''$ overlap. This involved taking many dis-registered short exposures of each 'quadrant' in a $3 \times 3$ pattern of step size $10''$. The individual exposures were 120 s, divided into $12 \times 10$ s background-limited sub-exposures. Since the fields are relatively crowded, blank nearby sky patches were also imaged for flat-fielding purposes. Dark frames were frequently obtained with the same exposure times and subtracted from the science images.

The median[†] of individual images (on source and on sky) taken over an $\simeq 18$ minute period around a given observation produced a flat-field, which, when applied, yielded images uniform to better than 1 part in $10^4$.

The large pixels did not fully sample the best seeing, but given images with dis-registered positions and fractional pixel shifts, information is present on smaller spatial scales. To take advantage of this, we artificially divided each pixel into four to obtain a pixel scale of $0.395''$ pixel$^{-1}$. After registering (using sub-pixel shifts), the images were median combined to give a well-sampled seeing of $1.1''$ and mosaicked to cover $170'' \times 170''$ at their maximum depth. Reductions were executed using the FIGARO package written by Keith Shortridge.

Photometric calibrations were secured by repeated observations of standard stars in the Allen (1992) list, yielding absolute photometry with an accuracy better than 0.03 mag *r.m.s.* Here we present the magnitudes in the $K'$ system. A transformation to the standard $K$ system can be achieved using $K = K' + 0.002 - 0.096(H - K)$; the colour term has been determined by numerically folding the filter response curves with the near infrared galaxy SED's described in Aragón-Salamanca et al. (1993, hereafter AECC). Non-evolving normal galaxy SED's typically have $(H - K') \simeq 0.75 \pm 0.05$ at $z = 0.31$, independent of galaxy type.

In order to construct $K'$–selected galaxy samples, automated object detection was performed using the APM software (Irwin 1985) in the STARLINK PISA implementation. An object was detected in the final $K'$ images when 4 or more connected pixels each had counts larger than $5\sigma$ above the background. Integrated photometry in a $5''$ diameter aperture was obtained using the PHOTOM package in STARLINK. This was preferred to using the APM magnitudes in that it allows for a local estimation of the sky.

The catalogues are complete to $K' \simeq 19$ mag, and the $5\sigma$ detection limit in the $5''$ aperture was $K' \simeq 20$ in all three cluster fields. Random photometric errors were determined empirically from the scatter estimated in sub-exposures and by using the sky variance near each object, since for the high $K$-band background this should be the dominant source of error. Both methods agree very well, yielding $\simeq 0.04$ at $K' \simeq 18$ and $\simeq 0.07$ at $K' \simeq 19$.

---

[†] Strictly speaking, instead of the median we used the Biweight central location estimator, which behaves like the median when the number of data points is large but produces a better S/N for relatively small samples. See Beers et al. (1990) for a detailed description of this statistical estimator.

**Table 2.** Log of the $K'$−band observations.

| Cluster | RA(1950) | Dec(1950) | Date | Seeing ($''$) | On Source Exposure (ks) |
|---------|----------|-----------|------|-------------|-------------------------|
| AC114 | $22^h 56^m 12^s$ | $-35°04'$ | Sep 1993 | 1.1 | 7.56 |
| AC118 | $00^h 11^m 48^s$ | $-30°42'$ | Sep 1993 | 1.7 | 7.56 |
| AC103 | $20^h 53^m 06^s$ | $-64°51'$ | Sep 1993 | 1.4 | 6.48 |

Although the catalogues are complete to $K' \simeq 19$, we will restrict our colour study to $K' \simeq 18$, since this limit approximately matches the depth of the ground-based and HST optical data. Only when discussing the luminosity function will we use the $K' \leq 19$ galaxy samples.

### 4.2.2  $K$−band cluster luminosity function

In order to fully exploit our infrared data for modelling cluster evolution we need to determine the form of the cluster $K$-band LF. This is also of interest in constraining possible luminosity evolution in the quiescent early-type galaxies. Although AECC found little evidence for any evolution in the $K$-band LF with earlier data, the data discussed above is more comprehensive, reaching much fainter absolute magnitudes with a much larger number of galaxies. Our combined cluster sample contains 484 galaxies to $K' = 19$. Field contamination estimated using published $K$ counts (Gardner, Cowie & Wainscoat 1994) suggests $\simeq 330$ are cluster members.

Figure 7 shows the combined cluster LF after field correction, assuming $H_0 = 50\,\mathrm{km\,s^{-1}\,Mpc^{-1}}$, $q_0 = 0.5$, and a $k$−correction of $-0.55$ mag, consistent with data discussed by AECC. The best-fitting Schechter function (Schechter 1976) (solid line) yields $M_{K'}^\star = -25.23^{+0.13}_{-0.14}$ and $\alpha = -1.0^{+0.13}_{-0.11}$. The former value is indistinguishable from the $M_{K'}^\star$ values determined by AECC for $0.37 \leq z \leq 0.92$ clusters.

Due to the present lack of a nearby cluster $K$−band LF, our results can only be compared with the $z \sim 0$ field LF of Mobasher, Sharples & Ellis (1993). They found no evidence for differences between the LF's of early-types and spirals. Their magnitudes were measured in an effective aperture $D_0$ (the galaxy diameter at $\mu_B = 25\,\mathrm{mag/arcsec^2}$). Our $5''$ aperture corresponds to a metric aperture of $\simeq 28$ kpc at $z = 0.31$, which was found to correspond very closely to the average $D_0$ of early-type galaxies in the Coma cluster (Bower, Lucey & Ellis 1992); thus, a direct comparison between our magnitudes and those of Mobasher et al. can safely be made. Their best estimate Schechter function parameters are $M_K^\star = -25.1 \pm 0.3$ and $\alpha = -1.0 \pm 0.3$, in close agreement with that determined above. This suggests no luminosity evolution in $K'$ since $z = 0.3$.

### 4.2.3  Optical and ultra-violet photometry

The optical and $U$−band CCD images were obtained on the CTIO 4.0-m and ESO NTT 3.5-m telescopes. A log of these observations is given in Table 3. The data were reduced in a standard manner using IRAF. Absolute photometric calibration was obtained via frequent observations of standard stars from Graham (1982). However, the conditions were such that we cannot rule out zeropoint uncertainties $\approx 0.2$ mag. Photometry was performed using apertures matching those used in the infrared. Error analysis similar to that described in §4.2.1 yielded the internal photometric errors presented in Tables 4 to 6. The tables contain the final $K'IU$ magnitudes (and $R$ for AC114) and the positions in RA and Dec (1950) for the $K' = 18$ limited samples (277 stars and galaxies in the 3 clusters). In the 49 cases where a $K'$−detected object was not seen in the $U$−band, we have given $\sim 3\sigma$ upper limits on $U$. Using published $K$−counts (Gardner et al. 1994), we expect the number of contaminating field galaxies to $K' = 18$ in the combined dataset covering the range $0 < U - I < 6$ to be 75.

Stellar contamination has been estimated by plotting Galactic stars on the $U - I$ versus $I - K'$ plane (Johnson 1966; Koornneef 1983) with the expected loci of galaxies of different spectral types at different redshifts. Regardless of their nature and redshift, galaxies are well separated from stars in this colour-colour plane, and hence the object colours are useful for stellar discrimination. We also independently checked this procedure by examining the HST and $K'$−band image profiles. Objects classified as stars are marked in the tables with asterisks and have been excluded from further analyses.

### 4.2.4  Colour-magnitude diagrams

Figures 8 show the $I - K'$ and $U - K'$ versus $K'$ colour-magnitude (C-M) diagrams for the three clusters. These include all non-stellar objects in the photometric catalogues. The expected location of the C-M sequence based on the Coma data of Bower et al. (1992) was calculated as described by AES. A red-envelope is clearly visible in all diagrams, and its location and slope is compatible with negligible colour evolution between $z \sim 0$ and $z = 0.31$ for the reddest galaxies, in agreement with AES. However, zeropoint uncertainties of $< 0.2$ mag noted in §4.2.3 preclude a detailed comparison. The importance of the rest-frame 2000 Å $U$−band in detecting recent star formation is illustrated by the size of the blueward scatter from the C-M sequence in the $U - K'$ figure.

Note the presence in the AC103 case of three relatively bright ($K' < 17$) objects with $I - K' < 2$ (objects 1, 68, and 193). Object 68 is a field spiral; the other two are border-line objects in the star-galaxy separation using colour criteria (see section 5.3). In the AC103 optical and infrared images object 1 appears to be a blend of several objects, possibly including a star, but it is not clear what object 193 is. Since AC103 is the cluster field with the highest stellar contamination, we suspect they are probably stars, but we have kept them for completeness.

**Table 3.** Log of the optical observations.

| Cluster | Telescope | Scale ($''/pixel$) | Date | Seeing ($''$) | Band | On Source Exposure (ks) |
|---------|-----------|--------------------|------|---------------|------|--------------------------|
| AC114 | 3.5m NTT | 0.351 | Aug 1993 | 1.4 | $R$ | 2.2 |
| | 4m CTIO | 0.432 | Sep 1993 | 1.5 | $I$ | 2.4 |
| | 3.5m NTT+4m CTIO | 0.362 | Aug/Sep 1993 | 1.4 | $U$ | 20.0 |
| AC118 | 4m CTIO | 0.432 | Sep 1993 | 1.1 | $I$ | 7.0 |
| | 3.5m NTT | 0.362 | Aug 1993 | 1.2 | $U$ | 2.0 |
| AC103 | 4m CTIO | 0.432 | Sep 1993 | 1.6 | $I$ | 1.0 |
| | 4m CTIO | 0.432 | Sep 1993 | 1.8 | $U$ | 6.0 |

In each C-M diagram there are a few faint objects with colours substantially redder than the C-M sequence. Without redshifts, we cannot rule out the possibility that these are high redshift field galaxies. However, since objects of such colour are rare in the field (Cowie et al. 1994), and several examples were found by AECC, we suspect they could be cluster members. If correct, their colours present a major challenge to current models of galaxy evolution, although possible explanations include exotic initial mass functions (Charlot et al. 1993) or very high metallicities (Glazebrook et al. 1995). Near-infrared spectroscopy may shed light on the puzzling nature of these galaxies.

*4.2.5 Colour-colour diagram*

In Figs. 9 we compare the $U - I$ versus $I - K'$ distribution for the combined $K' \le 18$ sample with the same model distribution which gave a reasonable fit in Fig. 5b. When compared to the observations, the models were found to be too blue in the $I - K'$ axis by $\sim 0.3$ mag, a well-known problem with the BC code when used to model galaxy infrared colours (Cole et al. 1994; Alonso-Herrero et al. 1995). We expect the accuracy of the models to be better in predicting relative changes rather than absolute values. For Fig. 9b we applied the offset to the models such that the quiescent model galaxies would coincide with the non-bursting population, whose colours were obtained from local galaxy spectra folded with the appropriate filters and observed at a redshift of 0.31.

The photometric data provide much larger samples than were available for the $H\delta$ versus $B - R$ comparison, although not all are cluster members. The model calculations were carried out as before, except this time magnitudes were assigned according to the Mobasher et al. (1993) local $K$-band LF corrected to $z = 0.31$. The colour plane was partitioned into quiescent ($U - I \ge 3.6$), blue ($2.5 \le U - I < 3.6$), and UV-strong ($U - I < 2.5$) regions, and we aimed to match the field-subtracted numbers $N_{quies.}$, $N_{blue}$, and $N_{UV}$.

Figures 9 show that, under the assumption of a secondary burst cycle the 30 per cent fraction derived earlier provides a consistent fit to the numbers (corrected for field contamination) observed in the UV-strong and blue regions of the colour-colour plane with similar uncertainties as before. The final column in Table 1 gives the average $\chi^2$ values over all the spectral and photometric categories. The 60:40 spheroidal:spiral mix with a 30 per cent bursting population minimizes the $\chi^2$ statistic for the observed versus predicted frequencies.

The additional check provided by the $U$-band data is valuable in three respects. Firstly, because the data are $K'$-selected, the comparison is hardly affected by uncertainties in luminosity boosting. Secondly, by virtue of the deep high quality infrared images and the small field contamination in the cluster cores, the comparison with the models is, in contrast to the spectroscopic samples, unaffected by small number statistics. Finally, as the UV light is particularly sensitive to the burst phase (c.f. Fig. 1c), we secure a valuable check on a combination of the duration of the burst and the mass fraction involved. The quiescent region of the diagram, dominated by non-bursting spheroidals, is also a good constraint on the relative proportion of spheroidal and spiral members.

One test of the single star formation cycle follows from the requirement that if the HDS phase follows sequentially from the SB and PSG phases, then we would expect the more evolved galaxies to be systematically fainter (c.f. Fig. 2). The argument is obviously statistical since, for a single observation, different galaxies are seen in the various phases. Unfortunately, examination of the current sample reveals a wide range of $K'$ luminosities for the various phases and, as the decline in luminosity with time for a model galaxy is only $\simeq 0.4 - 0.6$ mag (Fig. 2), very large samples would be needed to make this a decisive test. Nonetheless, it remains an important constraint for the future.

Thus, in summary, the active cluster fraction determined from adopting a single cycle of activity in the $H\delta - (B - R)$ plane can reproduce the density of statistically-determined members in the $(U - I) - (I - K')$ plane.

**Table 4.** AC114: aperture=5″; $K' \leq 18.0$ mag

| # | RA(1950) | Dec(1950) | $K'$ | $\sigma_{K'}$ | $I$ | $\sigma_I$ | $R$ | $\sigma_R$ | $U$ | $\sigma_U$ |
|---|----------|-----------|------|------|------|------|------|------|------|------|
| *131 | 22 55 54.80 | -35 03 37.6 | 13.05 | 0.00 | 15.01 | 0.00 | 14.80 | 0.00 | 17.16 | 0.00 |
| *180 | 22 56 01.65 | -35 02 46.3 | 13.23 | 0.00 | 15.65 | 0.00 | 16.96 | 0.00 | 20.92 | 0.01 |
| *132 | 22 55 55.01 | -35 03 31.3 | 13.45 | 0.00 | 15.11 | 0.00 | 14.93 | 0.00 | 17.14 | 0.00 |
| 88 | 22 56 02.05 | -35 04 13.8 | 14.61 | 0.01 | 17.46 | 0.01 | 18.15 | 0.01 | 21.58 | 0.02 |
| 115 | 22 55 55.70 | -35 03 51.8 | 14.80 | 0.01 | 17.55 | 0.00 | 18.23 | 0.01 | 21.58 | 0.02 |
| *150 | 22 56 01.47 | -35 03 16.5 | 14.81 | 0.01 | 16.74 | 0.01 | 17.93 | 0.00 | 22.01 | 0.03 |
| *142 | 22 56 06.20 | -35 03 26.9 | 14.83 | 0.01 | 16.70 | 0.01 | 17.62 | 0.01 | 21.21 | 0.01 |
| *52 | 22 55 55.43 | -35 04 35.9 | 15.06 | 0.01 | 17.16 | 0.00 | 18.57 | 0.01 | 22.50 | 0.04 |
| 94 | 22 56 01.28 | -35 04 20.1 | 15.19 | 0.01 | 17.92 | 0.01 | 18.60 | 0.01 | 21.72 | 0.03 |
| 13 | 22 55 58.22 | -35 05 16.5 | 15.46 | 0.01 | 18.12 | 0.01 | 18.75 | 0.01 | 21.87 | 0.03 |
| 84 | 22 56 04.57 | -35 04 07.3 | 15.52 | 0.01 | 18.22 | 0.00 | 18.91 | 0.01 | 22.25 | 0.03 |
| 104 | 22 55 54.29 | -35 03 57.7 | 15.54 | 0.01 | 18.22 | 0.01 | 18.88 | 0.01 | 22.11 | 0.03 |
| 138 | 22 56 01.63 | -35 03 31.4 | 15.56 | 0.01 | 18.34 | 0.01 | 19.02 | 0.01 | 22.35 | 0.04 |
| *57 | 22 55 52.96 | -35 04 26.7 | 15.57 | 0.01 | 17.58 | 0.00 | 18.86 | 0.01 | 22.64 | 0.05 |
| 95 | 22 56 01.45 | -35 04 16.4 | 15.61 | 0.01 | 18.27 | 0.02 | 18.94 | 0.02 | 22.24 | 0.04 |
| 178 | 22 55 54.50 | -35 02 48.6 | 15.64 | 0.01 | 18.26 | 0.00 | 18.93 | 0.01 | 21.78 | 0.03 |
| 159 | 22 56 02.75 | -35 03 07.6 | 15.74 | 0.01 | 18.43 | 0.01 | 19.06 | 0.01 | 22.10 | 0.03 |
| *66 | 22 56 00.60 | -35 04 20.9 | 15.74 | 0.01 | 16.90 | 0.00 | 17.36 | 0.00 | 19.32 | 0.01 |
| 90 | 22 56 00.29 | -35 04 04.7 | 15.77 | 0.01 | 18.47 | 0.01 | 19.16 | 0.01 | 22.09 | 0.03 |
| 97 | 22 56 02.18 | -35 04 18.1 | 15.90 | 0.01 | 18.59 | 0.01 | 19.23 | 0.02 | 22.19 | 0.04 |
| 148 | 22 55 53.19 | -35 03 20.1 | 15.95 | 0.01 | 18.67 | 0.01 | 19.37 | 0.01 | 22.58 | 0.04 |
| 160 | 22 56 03.02 | -35 03 06.5 | 15.96 | 0.01 | 18.49 | 0.01 | 19.08 | 0.01 | 21.13 | 0.01 |
| 134 | 22 55 59.98 | -35 03 34.6 | 16.03 | 0.01 | 18.66 | 0.01 | 19.29 | 0.01 | 22.29 | 0.03 |
| 79 | 22 55 55.29 | -35 04 11.6 | 16.04 | 0.01 | 18.65 | 0.01 | 19.35 | 0.01 | 22.78 | 0.05 |
| 12 | 22 56 04.20 | -35 05 17.1 | 16.07 | 0.01 | 18.75 | 0.01 | 19.43 | 0.01 | 22.65 | 0.05 |
| 48 | 22 56 00.64 | -35 04 36.6 | 16.07 | 0.01 | 18.80 | 0.01 | 19.50 | 0.01 | 22.87 | 0.05 |
| 177 | 22 56 00.03 | -35 02 48.7 | 16.10 | 0.01 | 18.58 | 0.01 | 19.18 | 0.01 | 21.01 | 0.01 |
| 99 | 22 56 03.62 | -35 04 02.4 | 16.14 | 0.01 | 18.75 | 0.01 | 19.45 | 0.01 | 22.74 | 0.05 |
| 187 | 22 55 53.62 | -35 02 39.5 | 16.16 | 0.01 | 18.64 | 0.01 | 19.34 | 0.01 | 22.59 | 0.04 |
| *38 | 22 56 07.67 | -35 04 50.5 | 16.23 | 0.01 | 17.99 | 0.00 | 18.99 | 0.01 | 22.61 | 0.04 |
| 179 | 22 56 03.60 | -35 02 47.5 | 16.27 | 0.01 | 18.73 | 0.01 | 19.38 | 0.01 | 22.31 | 0.04 |
| 167 | 22 55 53.84 | -35 02 58.0 | 16.30 | 0.01 | 18.75 | 0.01 | 19.40 | 0.01 | 22.21 | 0.03 |
| *93 | 22 56 01.21 | -35 04 13.3 | 16.33 | 0.04 | 18.75 | 0.04 | 20.31 | 0.08 | 23.49 | 0.14 |
| 96 | 22 56 01.54 | -35 04 14.0 | 16.35 | 0.03 | 18.99 | 0.03 | 19.63 | 0.03 | 22.62 | 0.12 |
| 193 | 22 55 57.67 | -35 02 33.8 | 16.44 | 0.01 | 18.96 | 0.01 | 19.65 | 0.01 | 22.88 | 0.05 |
| 110 | 22 56 00.56 | -35 03 55.1 | 16.47 | 0.01 | 19.09 | 0.01 | 19.74 | 0.01 | 22.68 | 0.05 |
| 188 | 22 56 00.00 | -35 02 38.1 | 16.52 | 0.01 | 18.74 | 0.01 | 19.33 | 0.01 | 21.70 | 0.02 |
| 89 | 22 56 00.72 | -35 04 07.0 | 16.54 | 0.02 | 19.13 | 0.02 | 19.81 | 0.02 | 23.03 | 0.09 |
| 111 | 22 56 00.21 | -35 03 53.0 | 16.54 | 0.01 | 19.21 | 0.01 | 19.89 | 0.01 | 23.19 | 0.08 |
| *176 | 22 55 57.56 | -35 02 48.9 | 16.61 | 0.01 | 18.62 | 0.01 | 19.91 | 0.01 | 22.69 | 0.05 |
| 139 | 22 55 58.92 | -35 03 32.2 | 16.65 | 0.01 | 19.21 | 0.01 | 19.90 | 0.01 | 23.11 | 0.06 |
| 77 | 22 55 59.66 | -35 04 14.7 | 16.69 | 0.01 | 19.26 | 0.01 | 19.92 | 0.01 | 22.83 | 0.06 |
| 163 | 22 56 02.59 | -35 03 02.3 | 16.73 | 0.01 | 19.17 | 0.01 | 19.81 | 0.01 | 22.69 | 0.05 |
| 162 | 22 56 05.19 | -35 03 03.4 | 16.89 | 0.01 | 19.14 | 0.01 | 19.70 | 0.01 | 22.08 | 0.03 |
| 175 | 22 55 55.33 | -35 02 51.4 | 16.91 | 0.02 | 19.14 | 0.01 | 19.77 | 0.01 | 22.60 | 0.04 |
| 49 | 22 56 00.31 | -35 04 37.4 | 16.91 | 0.01 | 21.16 | 0.07 | 22.00 | 0.07 | 25.00 | — |
| 145 | 22 55 57.90 | -35 03 15.21 | 16.91 | 0.01 | 19.52 | 0.01 | 20.18 | 0.02 | 23.44 | 0.08 |
| 114 | 22 55 53.42 | -35 03 54.8 | 16.99 | 0.02 | 19.51 | 0.02 | 20.12 | 0.02 | 23.42 | 0.09 |
| 183 | 22 55 56.67 | -35 02 44.5 | 17.03 | 0.02 | 19.44 | 0.01 | 20.13 | 0.02 | 23.03 | 0.06 |
| 135 | 22 56 00.28 | -35 03 35.0 | 17.09 | 0.02 | 19.44 | 0.01 | 19.94 | 0.02 | 21.44 | 0.02 |
| 181 | 22 56 04.14 | -35 02 44.6 | 17.12 | 0.02 | 19.56 | 0.02 | 20.11 | 0.02 | 22.76 | 0.07 |
| 166 | 22 55 56.64 | -35 02 59.0 | 17.12 | 0.02 | 20.14 | 0.02 | 21.00 | 0.03 | 23.58 | 0.09 |
| 171 | 22 55 59.29 | -35 02 57.0 | 17.13 | 0.02 | 19.66 | 0.02 | 20.32 | 0.02 | 23.71 | 0.10 |
| 23 | 22 56 02.81 | -35 05 07.9 | 17.16 | 0.02 | 19.75 | 0.01 | 20.41 | 0.02 | 23.59 | 0.09 |
| 123 | 22 55 55.94 | -35 03 45.8 | 17.17 | 0.02 | 19.76 | 0.03 | 20.38 | 0.03 | 23.76 | 0.15 |
| 5 | 22 56 01.05 | -35 05 34.0 | 17.21 | 0.02 | 19.84 | 0.02 | 20.67 | 0.02 | 22.75 | 0.05 |
| 51 | 22 55 59.51 | -35 04 37.2 | 17.23 | 0.02 | 20.33 | 0.03 | 21.63 | 0.05 | 24.62 | 0.21 |
| 68 | 22 55 57.94 | -35 04 20.7 | 17.26 | 0.02 | 19.72 | 0.02 | 20.39 | 0.02 | 23.45 | 0.08 |
| 182 | 22 56 02.17 | -35 02 45.1 | 17.29 | 0.02 | 19.85 | 0.02 | 20.61 | 0.03 | 23.87 | 0.12 |
| 4 | 22 56 04.46 | -35 05 34.4 | 17.32 | 0.02 | 20.45 | 0.03 | 21.69 | 0.05 | 23.34 | 0.08 |
| 87 | 22 56 01.66 | -35 04 25.0 | 17.35 | 0.04 | 19.96 | 0.05 | 20.50 | 0.04 | 23.81 | 0.12 |
| 67 | 22 55 58.89 | -35 04 21.4 | 17.36 | 0.02 | 19.85 | 0.02 | 20.60 | 0.02 | 23.88 | 0.13 |

**Table 4** – *continued*

| # | RA(1950) | Dec(1950) | $K'$ | $\sigma_{K'}$ | $I$ | $\sigma_I$ | $R$ | $\sigma_R$ | $U$ | $\sigma_U$ |
|---|----------|-----------|------|------|------|------|------|------|------|------|
| 31 | 22 56 05.77 | -35 04 55.6 | 17.37 | 0.02 | 19.85 | 0.02 | 20.50 | 0.02 | 23.84 | 0.11 |
| 18 | 22 56 00.25 | -35 05 14.9 | 17.39 | 0.02 | 19.87 | 0.02 | 20.57 | 0.02 | 23.57 | 0.09 |
| 37 | 22 56 06.55 | -35 04 51.5 | 17.39 | 0.02 | 19.36 | 0.01 | 19.87 | 0.01 | 21.08 | 0.01 |
| 129 | 22 55 52.60 | -35 03 41.7 | 17.40 | 0.02 | 19.36 | 0.01 | 20.65 | 0.03 | 25.00 | — |
| 44 | 22 56 06.38 | -35 04 42.5 | 17.41 | 0.02 | 19.77 | 0.02 | 20.44 | 0.02 | 23.27 | 0.07 |
| 141 | 22 56 02.67 | -35 03 29.7 | 17.43 | 0.02 | 20.18 | 0.02 | 20.87 | 0.03 | 24.18 | 0.14 |
| 72 | 22 56 03.64 | -35 04 18.6 | 17.44 | 0.02 | 19.48 | 0.01 | 19.87 | 0.01 | 21.36 | 0.02 |
| 92 | 22 56 01.28 | -35 04 08.0 | 17.45 | 0.09 | 20.03 | 0.10 | 20.94 | 0.11 | 25.00 | — |
| 33 | 22 55 56.78 | -35 04 53.1 | 17.47 | 0.02 | 19.58 | 0.01 | 20.01 | 0.02 | 21.11 | 0.01 |
| 2 | 22 55 55.51 | -35 05 36.3 | 17.50 | 0.02 | 20.33 | 0.03 | 21.31 | 0.04 | 24.24 | 0.16 |
| 91 | 22 56 00.99 | -35 04 08.5 | 17.53 | 0.05 | 20.10 | 0.07 | 20.85 | 0.06 | 23.79 | 0.17 |
| 105 | 22 56 03.25 | -35 04 00.3 | 17.53 | 0.02 | 20.10 | 0.03 | 20.70 | 0.03 | 23.52 | 0.10 |
| 151 | 22 56 02.35 | -35 03 16.7 | 17.54 | 0.02 | 20.05 | 0.02 | 20.72 | 0.03 | 23.62 | 0.09 |
| 119 | 22 56 01.38 | -35 03 51.6 | 17.57 | 0.02 | 20.19 | 0.03 | 20.92 | 0.03 | 24.34 | 0.17 |
| 22 | 22 55 59.33 | -35 05 09.1 | 17.57 | 0.02 | 20.17 | 0.02 | 20.75 | 0.03 | 22.41 | 0.04 |
| 161 | 22 56 04.68 | -35 03 03.8 | 17.57 | 0.02 | 19.55 | 0.02 | 19.88 | 0.01 | 20.90 | 0.01 |
| 116 | 22 55 55.27 | -35 03 47.0 | 17.64 | 0.03 | 20.32 | 0.05 | 21.01 | 0.05 | 22.31 | 0.04 |
| 82 | 22 55 57.16 | -35 04 09.3 | 17.66 | 0.02 | 19.17 | 0.01 | 19.41 | 0.01 | 19.78 | 0.01 |
| 64 | 22 55 58.11 | -35 04 24.1 | 17.75 | 0.02 | 20.21 | 0.03 | 20.86 | 0.03 | 24.41 | 0.19 |
| 169 | 22 55 55.47 | -35 02 58.5 | 17.76 | 0.03 | 20.15 | 0.03 | 20.85 | 0.03 | 24.12 | 0.18 |
| 125 | 22 55 53.43 | -35 03 44.1 | 17.78 | 0.03 | 21.12 | 0.06 | 21.79 | 0.06 | 23.86 | 0.13 |
| 75 | 22 56 00.42 | -35 04 17.0 | 17.80 | 0.05 | 20.16 | 0.05 | 20.64 | 0.04 | 23.27 | 0.11 |
| 112 | 22 55 57.49 | -35 03 55.8 | 17.81 | 0.02 | 20.30 | 0.03 | 20.98 | 0.03 | 22.66 | 0.05 |
| 63 | 22 56 00.08 | -35 04 25.3 | 17.81 | 0.03 | 20.34 | 0.04 | 21.06 | 0.04 | 22.93 | 0.07 |
| 144 | 22 56 01.04 | -35 03 26.0 | 17.89 | 0.03 | 20.15 | 0.03 | 20.62 | 0.03 | 22.97 | 0.06 |
| 14 | 22 56 05.05 | -35 05 17.4 | 17.91 | 0.03 | 20.65 | 0.05 | 21.35 | 0.05 | 24.44 | 0.21 |
| 7 | 22 56 01.91 | -35 05 28.9 | 17.91 | 0.03 | 20.11 | 0.02 | 20.83 | 0.03 | 22.87 | 0.05 |
| 172 | 22 56 03.44 | -35 02 57.0 | 17.92 | 0.03 | 20.75 | 0.04 | 21.67 | 0.05 | 25.00 | — |
| 122 | 22 56 01.21 | -35 03 47.8 | 17.93 | 0.03 | 20.45 | 0.03 | 21.14 | 0.04 | 24.40 | 0.17 |
| 152 | 22 56 03.19 | -35 03 15.0 | 17.98 | 0.03 | 20.33 | 0.04 | 21.00 | 0.03 | 23.78 | 0.11 |

## 5  COMPARISONS WITH DYNAMICAL SIMULATIONS AND HST MORPHOLOGIES

The detailed model comparisons above have illustrated how it is possible to reconcile the distribution of colours and spectral features of cluster members with an ongoing cycle of activity that is noticeably absent in the cores of present day clusters. The most significant conclusion that arises directly from our population studies is the large proportion ($\simeq 30$ per cent) of cluster members that must undergo this activity within $\sim 2$ Gyr of the time of observation in order to explain both the number of UV-bright and HDS objects. This is understandable given the extent of star-formation activity comprising the Butcher-Oemler effect must be greater than that revealed in a single snapshot of a cluster.

Of course, the modelling of stellar populations with artificially superimposed secondary bursts does not lead directly to the origin of the starburst activity. However, if only one mechanism predominantly produces the cycle of activity, then a major conclusion from our work is that it has to be widespread, probably affecting both spheroidal and spiral galaxies alike. The shorter the active phase, the greater the proportion of galaxies undergoing the cycle.

Couch et al. (1994) and Dressler et al. (1994) used their HST images to examine the morphological nature of those cluster galaxies in various categories, thereby adding a new dimension to the discussions of the preceding sections. Couch et al. analysed confirmed members in AC114 and Abell 370 and found a high proportion of SB and PSG galaxies were disk galaxies involved in interactions and/or mergers. Similar conclusions were derived by Dressler et al. using a larger sample, although many of the galaxies were not spectroscopically-confirmed members. Couch et al. found most of the HDS galaxies, in contrast, appeared to be isolated spheroidals. The latter conclusion was recently challenged by Wirth et al. (1994) who examined morphological-spectral correlations for red galaxies in 0016+16 ($z = 0.54$) and found some disk galaxies in this category. However, the terminology of 'E+A' galaxies was generally adopted in this work (see §1) and the faint spectra were of lower resolution than those discussed here. We have re-examined this point with the WFPC-2 images of AC118 and AC103 recently received; a more detailed analysis of these images is forthcoming (Couch et al, in preparation). Across all 3 clusters we now have 8 confirmed HDS members of which 5 are Es and 2, possibly 3, are S0s. By contrast, of 5 PSGs (a category which would be included in Wirth et al's 'E+A' category), 2-4 show signs of active merging, 1-2 are spirals and 1 is possibly an Irregular. The updated comparison tends to support the earlier claim of Couch et al. and indicates clearly the importance of distinguishing between the PSG and HDS classes via spectroscopy of good signal/noise.

In an interesting article, Mihos (1995) simulated the morphological evolution of merging galaxies viewed at various stages at a fixed redshift by both an aberrated and repaired HST. He categorised the morphological evolution in various stages: (i) tidal tails prior to the actual merger,

**Table 5.** AC118: aperture=5″; $K' \leq 18.0$ mag

| # | RA(1950) | Dec(1950) | $K'$ | $\sigma_{K'}$ | $I$ | $\sigma_I$ | $U$ | $\sigma_U$ |
|---|---|---|---|---|---|---|---|---|
| *173 | 00 11 44.70 | -30 38 43.4 | 11.24 | 0.00 | 15.16 | 0.00 | 14.43 | 0.00 |
| 39 | 00 11 50.49 | -30 41 01.1 | 14.90 | 0.01 | 17.81 | 0.00 | 22.30 | 0.08 |
| 71 | 00 11 49.12 | -30 40 41.1 | 14.93 | 0.01 | 17.76 | 0.01 | 22.02 | 0.06 |
| 142 | 00 11 41.39 | -30 39 15.8 | 14.95 | 0.01 | 17.75 | 0.00 | 22.24 | 0.07 |
| 76 | 00 11 48.89 | -30 40 20.2 | 15.40 | 0.01 | 18.11 | 0.00 | 22.20 | 0.07 |
| 193 | 00 11 43.02 | -30 38 29.3 | 15.47 | 0.01 | 17.99 | 0.00 | 21.76 | 0.05 |
| 170 | 00 11 49.82 | -30 38 49.3 | 15.48 | 0.01 | 17.82 | 0.00 | 20.88 | 0.03 |
| 40 | 00 11 50.61 | -30 40 57.2 | 15.53 | 0.01 | 18.35 | 0.01 | 22.50 | 0.09 |
| 129 | 00 11 41.40 | -30 39 36.2 | 15.57 | 0.01 | 18.34 | 0.00 | 22.55 | 0.09 |
| 86 | 00 11 51.16 | -30 40 10.5 | 15.58 | 0.01 | 18.12 | 0.00 | 22.00 | 0.06 |
| 178 | 00 11 43.98 | -30 38 40.8 | 15.59 | 0.01 | 18.14 | 0.03 | 22.22 | 0.15 |
| 23 | 00 11 47.54 | -30 41 13.6 | 15.71 | 0.01 | 18.44 | 0.01 | 22.40 | 0.08 |
| 181 | 00 11 44.09 | -30 38 36.8 | 15.76 | 0.01 | 18.21 | 0.01 | 22.10 | 0.11 |
| 161 | 00 11 43.35 | -30 38 58.4 | 15.81 | 0.01 | 18.23 | 0.01 | 21.02 | 0.03 |
| 177 | 00 11 45.29 | -30 38 48.5 | 15.81 | 0.01 | 18.22 | 0.01 | 22.05 | 0.08 |
| 131 | 00 11 54.72 | -30 39 36.2 | 15.86 | 0.01 | 18.60 | 0.00 | 21.98 | 0.06 |
| 180 | 00 11 44.36 | -30 38 49.1 | 15.89 | 0.01 | 18.05 | 0.01 | 21.31 | 0.05 |
| 75 | 00 11 49.45 | -30 40 27.6 | 15.93 | 0.01 | 18.54 | 0.01 | 22.72 | 0.11 |
| *36 | 00 11 45.11 | -30 40 58.8 | 15.96 | 0.01 | 18.03 | 0.00 | 22.99 | 0.12 |
| 146 | 00 11 42.77 | -30 39 20.7 | 15.98 | 0.01 | 18.63 | 0.00 | 20.91 | 0.03 |
| 182 | 00 11 49.16 | -30 38 43.3 | 16.06 | 0.01 | 18.66 | 0.00 | 21.99 | 0.06 |
| 99 | 00 11 52.43 | -30 40 03.5 | 16.13 | 0.01 | 18.71 | 0.00 | 21.45 | 0.04 |
| 164 | 00 11 49.39 | -30 38 57.3 | 16.16 | 0.01 | 18.59 | 0.00 | 21.44 | 0.04 |
| 68 | 00 11 54.98 | -30 40 24.7 | 16.16 | 0.01 | 18.32 | 0.00 | 18.83 | 0.01 |
| 42 | 00 11 47.25 | -30 40 52.6 | 16.19 | 0.01 | 18.76 | 0.01 | 21.70 | 0.05 |
| 126 | 00 11 46.22 | -30 39 41.6 | 16.23 | 0.01 | 18.84 | 0.01 | 21.83 | 0.05 |
| 51 | 00 11 50.09 | -30 40 42.0 | 16.27 | 0.01 | 19.10 | 0.01 | 21.86 | 0.05 |
| 60 | 00 11 46.39 | -30 40 34.7 | 16.32 | 0.01 | 19.03 | 0.01 | 23.12 | 0.17 |
| 197 | 00 11 46.52 | -30 38 29.7 | 16.32 | 0.01 | 18.76 | 0.00 | 22.33 | 0.07 |
| 106 | 00 11 44.33 | -30 39 58.3 | 16.35 | 0.01 | 18.90 | 0.01 | 22.76 | 0.12 |
| 151 | 00 11 46.41 | -30 39 18.1 | 16.36 | 0.01 | 19.09 | 0.00 | 23.28 | 0.17 |
| *3 | 00 11 47.91 | -30 41 32.9 | 16.38 | 0.01 | 18.37 | 0.00 | 22.99 | 0.13 |
| 81 | 00 11 49.69 | -30 40 26.5 | 16.39 | 0.02 | 18.92 | 0.01 | 22.79 | 0.12 |
| 80 | 00 11 49.58 | -30 40 23.0 | 16.41 | 0.02 | 19.03 | 0.02 | 23.46 | 0.23 |
| 97 | 00 11 49.84 | -30 40 03.8 | 16.42 | 0.01 | 19.01 | 0.00 | 23.25 | 0.18 |
| 113 | 00 11 45.23 | -30 39 51.8 | 16.47 | 0.01 | 19.14 | 0.01 | 23.01 | 0.14 |
| 118 | 00 11 50.79 | -30 39 44.4 | 16.50 | 0.01 | 19.15 | 0.01 | 21.44 | 0.04 |
| 57 | 00 11 50.73 | -30 40 38.2 | 16.52 | 0.01 | 19.09 | 0.00 | 23.41 | 0.19 |
| 183 | 00 11 48.96 | -30 38 41.6 | 16.53 | 0.01 | 19.22 | 0.01 | 23.56 | 0.26 |
| 15 | 00 11 42.10 | -30 41 19.4 | 16.59 | 0.01 | 19.18 | 0.01 | 22.70 | 0.11 |
| 109 | 00 11 49.81 | -30 39 55.2 | 16.60 | 0.01 | 19.35 | 0.01 | 23.66 | 0.23 |
| 175 | 00 11 45.45 | -30 38 38.4 | 16.65 | 0.02 | 18.84 | 0.03 | 23.07 | 0.20 |
| 25 | 00 11 46.15 | -30 41 12.4 | 16.66 | 0.01 | 19.26 | 0.00 | 23.30 | 0.18 |
| 73 | 00 11 49.62 | -30 40 34.7 | 16.72 | 0.03 | 19.44 | 0.04 | 24.00 | — |
| 77 | 00 11 49.39 | -30 40 16.5 | 16.73 | 0.02 | 19.59 | 0.02 | 22.46 | 0.08 |
| 179 | 00 11 44.19 | -30 38 43.6 | 16.76 | 0.07 | 18.80 | 0.05 | 22.02 | 0.13 |
| 59 | 00 11 48.13 | -30 40 35.5 | 16.76 | 0.02 | 19.38 | 0.01 | 22.99 | 0.14 |
| 121 | 00 11 53.69 | -30 39 46.0 | 16.77 | 0.01 | 19.46 | 0.00 | 23.33 | 0.18 |
| 53 | 00 11 53.47 | -30 40 41.0 | 16.78 | 0.01 | 19.49 | 0.01 | 24.00 | — |
| 149 | 00 11 52.91 | -30 39 20.6 | 16.81 | 0.01 | 19.20 | 0.00 | 22.19 | 0.06 |
| 145 | 00 11 41.28 | -30 39 21.4 | 16.83 | 0.02 | 19.38 | 0.02 | 24.00 | — |
| 63 | 00 11 52.16 | -30 40 31.6 | 16.86 | 0.01 | 19.47 | 0.01 | 23.64 | 0.23 |
| *47 | 00 11 50.21 | -30 40 49.5 | 16.86 | 0.02 | 18.76 | 0.01 | 23.09 | 0.15 |
| 185 | 00 11 54.02 | -30 38 41.7 | 16.90 | 0.01 | 19.39 | 0.00 | 23.27 | 0.17 |
| 79 | 00 11 49.51 | -30 40 43.6 | 16.90 | 0.03 | 19.46 | 0.02 | 22.65 | 0.12 |
| 48 | 00 11 43.40 | -30 40 49.3 | 16.92 | 0.01 | 19.56 | 0.01 | 23.87 | 0.28 |
| 171 | 00 11 46.78 | -30 38 48.1 | 16.93 | 0.01 | 19.47 | 0.01 | 23.31 | 0.17 |
| 111 | 00 11 41.91 | -30 39 53.7 | 16.94 | 0.01 | 19.41 | 0.00 | 23.22 | 0.15 |
| 155 | 00 11 42.91 | -30 39 11.4 | 16.94 | 0.02 | 19.54 | 0.01 | 24.00 | — |
| 108 | 00 11 48.88 | -30 39 55.8 | 17.00 | 0.01 | 19.27 | 0.00 | 22.53 | 0.09 |
| 95 | 00 11 47.84 | -30 40 07.5 | 17.06 | 0.02 | 19.50 | 0.01 | 20.84 | 0.03 |
| 100 | 00 11 51.14 | -30 40 01.4 | 17.06 | 0.02 | 19.63 | 0.01 | 23.19 | 0.16 |

**Table 5** – *continued*

| # | RA(1950) | Dec(1950) | $K'$ | $\sigma_{K'}$ | $I$ | $\sigma_I$ | $U$ | $\sigma_U$ |
|---|----------|-----------|------|------|------|------|------|------|
| 56 | 00 11 41.07 | -30 40 39.9 | 17.09 | 0.02 | 19.80 | 0.01 | 22.02 | 0.06 |
| 144 | 00 11 41.21 | -30 39 09.0 | 17.10 | 0.03 | 19.64 | 0.02 | 23.77 | 0.27 |
| 87 | 00 11 51.02 | -30 40 06.4 | 17.13 | 0.02 | 19.58 | 0.02 | 23.75 | 0.30 |
| 163 | 00 11 41.75 | -30 38 57.6 | 17.13 | 0.02 | 19.68 | 0.01 | 23.49 | 0.20 |
| 128 | 00 11 44.06 | -30 39 39.2 | 17.15 | 0.02 | 19.72 | 0.01 | 23.87 | 0.30 |
| 110 | 00 11 44.12 | -30 39 54.1 | 17.16 | 0.02 | 19.57 | 0.01 | 23.14 | 0.17 |
| 52 | 00 11 53.69 | -30 40 43.4 | 17.17 | 0.02 | 19.82 | 0.01 | 23.46 | 0.19 |
| 33 | 00 11 51.43 | -30 41 03.0 | 17.17 | 0.02 | 19.89 | 0.02 | 23.21 | 0.16 |
| 32 | 00 11 43.76 | -30 41 03.3 | 17.19 | 0.02 | 19.75 | 0.01 | 21.42 | 0.04 |
| 4 | 00 11 53.45 | -30 41 32.8 | 17.21 | 0.02 | 19.74 | 0.01 | 23.75 | 0.25 |
| 162 | 00 11 43.73 | -30 38 59.4 | 17.22 | 0.02 | 19.78 | 0.04 | 23.48 | 0.24 |
| 94 | 00 11 50.09 | -30 40 08.0 | 17.24 | 0.02 | 19.77 | 0.01 | 23.69 | 0.28 |
| 148 | 00 11 42.10 | -30 39 20.9 | 17.25 | 0.03 | 19.90 | 0.02 | 23.53 | 0.21 |
| 43 | 00 11 54.12 | -30 40 53.6 | 17.27 | 0.02 | 19.02 | 0.00 | 21.08 | 0.03 |
| 38 | 00 11 50.42 | -30 41 08.6 | 17.28 | 0.03 | 19.94 | 0.02 | 24.00 | — |
| 19 | 00 11 42.57 | -30 41 18.9 | 17.30 | 0.03 | 19.85 | 0.01 | 23.30 | 0.18 |
| 96 | 00 11 45.56 | -30° 40 06.1 | 17.30 | 0.02 | 19.66 | 0.01 | 23.37 | 0.18 |
| 120 | 00 11 49.57 | -30 39 46.6 | 17.31 | 0.02 | 19.82 | 0.01 | 23.72 | 0.26 |
| *158 | 00 11 44.59 | -30 39 05.4 | 17.33 | 0.02 | 16.95 | 0.00 | 22.70 | 0.11 |
| 115 | 00 11 45.91 | -30 39 47.7 | 17.33 | 0.02 | 20.45 | 0.02 | 23.81 | 0.29 |
| 114 | 00 11 46.06 | -30 39 50.5 | 17.34 | 0.01 | 19.69 | 0.01 | 22.88 | 0.12 |
| 153 | 00 11 47.98 | -30 39 16.0 | 17.38 | 0.02 | 19.76 | 0.01 | 23.80 | 0.26 |
| 65 | 00 11 47.35 | -30 40 30.1 | 17.39 | 0.02 | 19.94 | 0.01 | 23.86 | 0.28 |
| 117 | 00 11 50.62 | -30 39 47.7 | 17.39 | 0.02 | 20.09 | 0.01 | 23.58 | 0.24 |
| 101 | 00 11 51.33 | -30 39 59.9 | 17.42 | 0.02 | 19.83 | 0.01 | 24.00 | — |
| 191 | 00 11 42.11 | -30 38 35.5 | 17.50 | 0.02 | 19.72 | 0.01 | 23.06 | 0.14 |
| 78 | 00 11 48.81 | -30 40 15.1 | 17.53 | 0.02 | 20.05 | 0.01 | 24.00 | — |
| 2 | 00 11 52.75 | -30 41 35.8 | 17.57 | 0.02 | 19.93 | 0.01 | 23.70 | 0.23 |
| 135 | 00 11 51.30 | -30 39 30.1 | 17.58 | 0.02 | 19.93 | 0.01 | 24.00 | — |
| 174 | 00 11 43.71 | -30 38 43.9 | 17.59 | 0.02 | 19.70 | 0.02 | 22.07 | 0.07 |
| 26 | 00 11 52.06 | -30 41 10.8 | 17.59 | 0.02 | 19.33 | 0.01 | 20.92 | 0.03 |
| 84 | 00 11 47.95 | -30 40 18.8 | 17.60 | 0.02 | 20.12 | 0.01 | 23.89 | 0.29 |
| 21 | 00 11 49.82 | -30 41 15.2 | 17.61 | 0.02 | 20.10 | 0.01 | 23.42 | 0.19 |
| 137 | 00 11 46.89 | -30 39 26.6 | 17.64 | 0.02 | 20.42 | 0.01 | 23.20 | 0.16 |
| 122 | 00 11 41.69 | -30 39 45.2 | 17.64 | 0.03 | 20.04 | 0.01 | 24.00 | — |
| 20 | 00 11 49.74 | -30 41 19.1 | 17.69 | 0.02 | 20.28 | 0.01 | 24.00 | — |
| 72 | 00 11 48.57 | -30 40 38.0 | 17.72 | 0.08 | 20.12 | 0.03 | 24.00 | — |
| 194 | 00 11 40.61 | -30 38 29.9 | 17.74 | 0.03 | 20.30 | 0.01 | 22.21 | 0.07 |
| 30 | 00 11 43.35 | -30 41 06.7 | 17.75 | 0.03 | 20.11 | 0.01 | 23.48 | 0.23 |
| 89 | 00 11 41.45 | -30 40 11.2 | 17.76 | 0.03 | 20.27 | 0.01 | 24.00 | — |
| 83 | 00 11 47.36 | -30 40 19.4 | 17.77 | 0.03 | 20.16 | 0.01 | 24.00 | — |
| 55 | 00 11 51.08 | -30 40 41.8 | 17.77 | 0.03 | 20.38 | 0.01 | 23.37 | 0.19 |
| 138 | 00 11 42.95 | -30 39 25.7 | 17.81 | 0.03 | 20.04 | 0.01 | 23.41 | 0.21 |
| 107 | 00 11 52.12 | -30 39 57.4 | 17.82 | 0.03 | 20.20 | 0.01 | 23.68 | 0.25 |
| 70 | 00 11 49.57 | -30 40 47.7 | 17.87 | 0.05 | 20.47 | 0.03 | 24.00 | — |
| 123 | 00 11 48.50 | -30 39 45.4 | 17.88 | 0.03 | 20.04 | 0.01 | 23.24 | 0.18 |
| 29 | 00 11 43.40 | -30 41 09.2 | 17.88 | 0.03 | 20.00 | 0.01 | 23.21 | 0.17 |
| *176 | 00 11 44.74 | -30 38 31.4 | 17.91 | 0.06 | 15.73 | 0.00 | 24.00 | — |
| *150 | 00 11 44.54 | -30 39 19.1 | 17.92 | 0.03 | 18.59 | 0.00 | 23.45 | 0.20 |
| 37 | 00 11 49.31 | -30 40 58.4 | 17.94 | 0.04 | 20.58 | 0.03 | 24.00 | — |
| 143 | 00 11 40.77 | -30 39 13.4 | 17.94 | 0.05 | 20.59 | 0.02 | 23.04 | 0.14 |
| 130 | 00 11 50.07 | -30 39 38.3 | 17.95 | 0.03 | 20.41 | 0.01 | 24.00 | — |
| 44 | 00 11 48.82 | -30 40 52.5 | 17.97 | 0.05 | 20.32 | 0.03 | 24.00 | — |
| 31 | 00 11 50.08 | -30 41 07.0 | 17.98 | 0.06 | 20.52 | 0.05 | 24.00 | — |
| 192 | 00 11 50.72 | -30 38 32.9 | 17.98 | 0.03 | 20.45 | 0.01 | 24.00 | — |

(ii) actual merging event with connected isophotes, (iii) faint tidal features surrounding a remnant, and (iv) a spheroidal remnant indistinguishable from a regular E/S0. Crude timescales are attached to each stage. Although simulation parameters, viewing angles, and other imponderables obviously affect the detailed numbers, Mihos makes two points. Firstly, stages (i)-(ii), as recognised by Couch et al. (1994), will be relatively brief ($< 1$ Gyr) in most encounters so the detection of *any* cases would imply a high frequency of merging. Secondly, it is not surprising in the merger picture that the HDS galaxies, which probe 1-2 Gyr after the burst, appear as regular spheroidals. In this case

Table 6. AC103: aperture=5″; $K' \leq 18.0$ mag

| # | RA(1950) | Dec(1950) | $K'$ | $\sigma_{K'}$ | $I$ | $\sigma_I$ | $U$ | $\sigma_U$ |
|---|---|---|---|---|---|---|---|---|
| *207 | 20 52 54.82 | -64 49 19.83 | 12.10 | 0.00 | 14.12 | 0.00 | 14.39 | 0.00 |
| *192 | 20 53 04.02 | -64 49 41.8 | 12.33 | 0.00 | 14.26 | 0.00 | 15.88 | 0.00 |
| *19 | 20 52 58.96 | -64 52 08.36 | 12.48 | 0.00 | 14.60 | 0.00 | 19.50 | 0.01 |
| *13 | 20 52 51.63 | -64 52 10.77 | 13.45 | 0.00 | 14.70 | 0.00 | 17.63 | 0.00 |
| *211 | 20 52 39.34 | -64 49 18.06 | 13.64 | 0.00 | 14.67 | 0.00 | 16.24 | 0.00 |
| *147 | 20 52 54.08 | -64 50 13.71 | 14.14 | 0.00 | 15.02 | 0.00 | 16.76 | 0.00 |
| *173 | 20 52 49.40 | -64 49 48.92 | 14.28 | 0.00 | 15.58 | 0.00 | 18.59 | 0.01 |
| *137 | 20 52 56.10 | -64 50 21.18 | 14.41 | 0.00 | 16.18 | 0.00 | 21.16 | 0.03 |
| 1 | 20 53 03.99 | -64 52 28.1 | 14.55 | 0.01 | 16.51 | 0.01 | 19.92 | 0.02 |
| *132 | 20 53 06.18 | -64 50 25.48 | 14.66 | 0.01 | 16.18 | 0.00 | 20.67 | 0.02 |
| 62 | 20 52 44.83 | -64 51 22.0 | 14.68 | 0.01 | 17.53 | 0.01 | 21.91 | 0.05 |
| 190 | 20 53 02.80 | -64 49 52.3 | 14.93 | 0.01 | 17.68 | 0.01 | 22.04 | 0.07 |
| *74 | 20 53 07.35 | -64 51 22.17 | 14.98 | 0.01 | 16.00 | 0.00 | 18.53 | 0.01 |
| *210 | 20 52 54.16 | -64 49 19.06 | 15.07 | 0.01 | 15.90 | 0.00 | 16.50 | 0.00 |
| 83 | 20 53 01.72 | -64 51 14.8 | 15.19 | 0.01 | 18.02 | 0.01 | 22.54 | 0.06 |
| 45 | 20 52 38.85 | -64 51 43.1 | 15.21 | 0.01 | 17.77 | 0.01 | 21.60 | 0.03 |
| *145 | 20 52 57.95 | -64 50 16.1 | 15.27 | 0.01 | 15.81 | 0.00 | 17.01 | 0.00 |
| 142 | 20 52 59.70 | -64 50 16.5 | 15.34 | 0.01 | 17.84 | 0.01 | 21.23 | 0.03 |
| 146 | 20 53 07.73 | -64 50 14.6 | 15.36 | 0.01 | 17.88 | 0.01 | 21.38 | 0.03 |
| *2 | 20 53 03.28 | -64 52 26.1 | 15.37 | 0.01 | 16.97 | 0.01 | 20.59 | 0.04 |
| 70 | 20 52 44.79 | -64 51 26.9 | 15.41 | 0.01 | 18.09 | 0.01 | 20.20 | 0.01 |
| *209 | 20 52 55.44 | -64 49 21.80 | 15.45 | 0.01 | 16.28 | 0.00 | 18.72 | 0.01 |
| *51 | 20 52 50.88 | -64 51 39.3 | 15.49 | 0.01 | 16.73 | 0.01 | 20.00 | 0.01 |
| 28 | 20 52 43.59 | -64 51 56.5 | 15.52 | 0.01 | 18.22 | 0.01 | 22.15 | 0.05 |
| 191 | 20 53 01.77 | -64 49 46.9 | 15.52 | 0.01 | 18.28 | 0.01 | 22.65 | 0.11 |
| 75 | 20 52 55.22 | -64 51 21.3 | 15.59 | 0.01 | 18.38 | 0.01 | 22.60 | 0.07 |
| 68 | 20 52 43.39 | -64 51 12.2 | 15.72 | 0.01 | 17.65 | 0.01 | 19.71 | 0.01 |
| 163 | 20 52 51.04 | -64 50 01.7 | 15.73 | 0.01 | 18.37 | 0.01 | 21.56 | 0.03 |
| *119 | 20 53 01.65 | -64 50 42.06 | 15.73 | 0.01 | 16.39 | 0.00 | 18.03 | 0.00 |
| 156 | 20 53 01.69 | -64 50 06.1 | 15.81 | 0.01 | 18.53 | 0.01 | 21.87 | 0.04 |
| *80 | 20 52 52.57 | -64 51 17.9 | 15.89 | 0.01 | 18.09 | 0.01 | 24.00 | — |
| *151 | 20 53 06.44 | -64 50 11.1 | 16.01 | 0.01 | 16.67 | 0.00 | 18.43 | 0.01 |
| 194 | 20 53 02.62 | -64 49 37.2 | 16.03 | 0.01 | 18.52 | 0.02 | 22.09 | 0.06 |
| 112 | 20 53 02.53 | -64 50 50.1 | 16.07 | 0.01 | 18.62 | 0.01 | 23.05 | 0.11 |
| 76 | 20 52 49.14 | -64 51 17.1 | 16.09 | 0.01 | 18.54 | 0.01 | 21.83 | 0.04 |
| 92 | 20 52 46.32 | -64 51 04.3 | 16.12 | 0.01 | 18.87 | 0.01 | 23.35 | 0.11 |
| 120 | 20 53 01.90 | -64 50 38.5 | 16.15 | 0.01 | 18.44 | 0.01 | 20.41 | 0.02 |
| 6 | 20 52 42.80 | -64 52 23.6 | 16.18 | 0.01 | 18.78 | 0.01 | 23.23 | 0.12 |
| 77 | 20 52 49.60 | -64 51 21.1 | 16.19 | 0.01 | 18.54 | 0.01 | 21.74 | 0.04 |
| 63 | 20 52 42.43 | -64 51 24.0 | 16.25 | 0.01 | 19.01 | 0.02 | 23.23 | 0.15 |
| 39 | 20 52 48.00 | -64 51 48.2 | 16.27 | 0.01 | 18.93 | 0.01 | 23.41 | 0.18 |
| *124 | 20 52 39.51 | -64 50 39.2 | 16.27 | 0.01 | 17.10 | 0.00 | 19.00 | 0.01 |
| 86 | 20 53 03.97 | -64 51 12.5 | 16.27 | 0.01 | 18.91 | 0.01 | 23.30 | 0.12 |
| 65 | 20 52 43.05 | -64 51 18.6 | 16.29 | 0.02 | 19.01 | 0.03 | 22.21 | 0.06 |
| 46 | 20 52 38.41 | -64 51 45.4 | 16.29 | 0.02 | 18.74 | 0.01 | 23.05 | 0.16 |
| 30 | 20 52 44.02 | -64 51 58.9 | 16.33 | 0.01 | 18.65 | 0.01 | 22.44 | 0.07 |
| 64 | 20 52 45.87 | -64 51 22.5 | 16.33 | 0.01 | 18.98 | 0.02 | 23.20 | 0.16 |
| 41 | 20 52 46.88 | -64 51 48.3 | 16.34 | 0.01 | 18.90 | 0.01 | 23.27 | 0.13 |
| 67 | 20 52 44.87 | -64 51 13.3 | 16.41 | 0.01 | 18.79 | 0.02 | 22.65 | 0.10 |
| 54 | 20 53 01.09 | -64 51 36.5 | 16.42 | 0.01 | 19.09 | 0.01 | 23.20 | 0.12 |
| 40 | 20 52 48.28 | -64 51 43.7 | 16.47 | 0.01 | 19.05 | 0.02 | 22.41 | 0.06 |
| *81 | 20 52 53.33 | -64 51 16.7 | 16.49 | 0.01 | 18.01 | 0.00 | 22.17 | 0.05 |
| *208 | 20 52 54.96 | -64 49 25.81 | 16.51 | 0.01 | 14.83 | 0.00 | 18.67 | 0.01 |
| *160 | 20 52 49.19 | -64 50 04.6 | 16.57 | 0.01 | 18.69 | 0.01 | 24.00 | — |
| 106 | 20 52 56.45 | -64 50 56.8 | 16.57 | 0.01 | 19.10 | 0.01 | 24.00 | — |
| *24 | 20 52 53.97 | -64 52 06.0 | 16.60 | 0.01 | 18.66 | 0.01 | 23.93 | 0.21 |
| 196 | 20 53 02.62 | -64 49 46.3 | 16.64 | 0.04 | 19.51 | 0.12 | 23.03 | 0.19 |
| *123 | 20 52 59.17 | -64 50 41.2 | 16.64 | 0.01 | 18.38 | 0.01 | 23.37 | 0.15 |
| *126 | 20 52 43.70 | -64 50 37.8 | 16.65 | 0.01 | 18.17 | 0.01 | 22.45 | 0.06 |
| *3 | 20 52 46.38 | -64 52 26.0 | 16.73 | 0.01 | 15.61 | 0.00 | 20.14 | 0.01 |
| 193 | 20 53 03.17 | -64 49 39.7 | 16.75 | 0.02 | 18.36 | 0.04 | 20.16 | 0.02 |
| 212 | 20 52 40.23 | -64 49 19.7 | 16.77 | 0.01 | 18.95 | 0.02 | 21.25 | 0.03 |

**Table 6** – *continued*

| # | RA(1950) | | | Dec(1950) | | | $K'$ | $\sigma_{K'}$ | $I$ | $\sigma_I$ | $U$ | $\sigma_U$ |
|---|---|---|---|---|---|---|---|---|---|---|---|---|
| 113 | 20 | 53 | 05.47 | -64 | 50 | 51.2 | 16.80 | 0.01 | 19.47 | 0.02 | 23.51 | 0.15 |
| 50 | 20 | 53 | 02.69 | -64 | 51 | 40.7 | 16.81 | 0.01 | 19.15 | 0.01 | 22.01 | 0.04 |
| *96 | 20 | 52 | 56.04 | -64 | 51 | 03.5 | 16.89 | 0.02 | 18.83 | 0.01 | 24.10 | 0.25 |
| 214 | 20 | 52 | 47.95 | -64 | 49 | 18.2 | 16.89 | 0.02 | 19.57 | 0.02 | 24.00 | — |
| 216 | 20 | 53 | 04.88 | -64 | 49 | 19.1 | 16.92 | 0.01 | 19.12 | 0.01 | 23.13 | 0.11 |
| 185 | 20 | 53 | 01.72 | -64 | 49 | 38.5 | 16.96 | 0.02 | 19.59 | 0.04 | 24.00 | — |
| 10 | 20 | 52 | 42.46 | -64 | 52 | 18.5 | 16.96 | 0.02 | 19.43 | 0.02 | 22.82 | 0.09 |
| 139 | 20 | 53 | 01.99 | -64 | 50 | 22.0 | 16.98 | 0.01 | 19.06 | 0.01 | 21.03 | 0.02 |
| 118 | 20 | 52 | 58.29 | -64 | 50 | 47.1 | 16.98 | 0.01 | 19.75 | 0.02 | 23.21 | 0.13 |
| 69 | 20 | 52 | 44.53 | -64 | 51 | 14.0 | 16.98 | 0.03 | 19.68 | 0.07 | 23.03 | 0.17 |
| 49 | 20 | 52 | 58.20 | -64 | 51 | 43.2 | 17.04 | 0.02 | 19.66 | 0.02 | 23.35 | 0.14 |
| 187 | 20 | 52 | 56.99 | -64 | 49 | 35.8 | 17.07 | 0.02 | 19.70 | 0.02 | 23.83 | 0.20 |
| 87 | 20 | 52 | 54.46 | -64 | 51 | 11.4 | 17.11 | 0.02 | 20.03 | 0.03 | 24.11 | 0.24 |
| 213 | 20 | 53 | 02.48 | -64 | 49 | 22.2 | 17.12 | 0.02 | 19.71 | 0.02 | 24.00 | — |
| 108 | 20 | 52 | 49.13 | -64 | 50 | 54.5 | 17.14 | 0.02 | 19.15 | 0.01 | 21.46 | 0.03 |
| 179 | 20 | 52 | 43.93 | -64 | 49 | 47.3 | 17.14 | 0.02 | 19.65 | 0.02 | 24.00 | — |
| 195 | 20 | 53 | 02.07 | -64 | 49 | 34.0 | 17.15 | 0.02 | 19.40 | 0.03 | 21.71 | 0.05 |
| 131 | 20 | 52 | 53.46 | -64 | 50 | 28.0 | 17.16 | 0.02 | 19.55 | 0.02 | 24.00 | — |
| 138 | 20 | 52 | 56.96 | -64 | 50 | 20.0 | 17.18 | 0.02 | 19.28 | 0.03 | 21.69 | 0.05 |
| 37 | 20 | 52 | 42.32 | -64 | 51 | 49.4 | 17.18 | 0.02 | 19.59 | 0.02 | 23.79 | 0.21 |
| 34 | 20 | 52 | 46.04 | -64 | 51 | 55.6 | 17.20 | 0.02 | 19.78 | 0.03 | 24.00 | — |
| 84 | 20 | 53 | 07.19 | -64 | 51 | 14.9 | 17.22 | 0.02 | 19.68 | 0.03 | 24.00 | — |
| 20 | 20 | 52 | 58.07 | -64 | 52 | 09.4 | 17.28 | 0.02 | 19.20 | 0.02 | 23.05 | 0.17 |
| 18 | 20 | 52 | 38.99 | -64 | 52 | 08.3 | 17.31 | 0.02 | 19.81 | 0.02 | 24.00 | — |
| *26 | 20 | 52 | 51.51 | -64 | 52 | 01.4 | 17.31 | 0.02 | 19.28 | 0.02 | 24.00 | — |
| 79 | 20 | 52 | 46.69 | -64 | 51 | 18.3 | 17.33 | 0.03 | 19.99 | 0.03 | 24.00 | — |
| 95 | 20 | 52 | 38.11 | -64 | 51 | 04.4 | 17.34 | 0.02 | 19.68 | 0.04 | 23.91 | 0.22 |
| 56 | 20 | 52 | 43.15 | -64 | 51 | 30.6 | 17.34 | 0.03 | 19.94 | 0.03 | 24.00 | — |
| 154 | 20 | 53 | 00.62 | -64 | 50 | 08.9 | 17.35 | 0.03 | 20.19 | 0.05 | 24.00 | — |
| 155 | 20 | 52 | 53.57 | -64 | 50 | 08.9 | 17.36 | 0.03 | 20.89 | 0.07 | 21.65 | 0.04 |
| *7 | 20 | 52 | 51.94 | -64 | 52 | 24.2 | 17.37 | 0.02 | 19.00 | 0.01 | 23.19 | 0.11 |
| *186 | 20 | 52 | 51.41 | -64 | 49 | 36.7 | 17.38 | 0.02 | 18.23 | 0.01 | 19.94 | 0.01 |
| *183 | 20 | 53 | 00.37 | -64 | 49 | 42.6 | 17.39 | 0.02 | 18.95 | 0.01 | 22.37 | 0.06 |
| 197 | 20 | 53 | 07.67 | -64 | 49 | 33.5 | 17.41 | 0.02 | 19.86 | 0.02 | 23.18 | 0.11 |
| 150 | 20 | 52 | 50.43 | -64 | 50 | 12.5 | 17.42 | 0.02 | 19.78 | 0.03 | 23.26 | 0.14 |
| 125 | 20 | 52 | 49.00 | -64 | 50 | 39.1 | 17.45 | 0.02 | 19.79 | 0.03 | 24.00 | — |
| 66 | 20 | 52 | 43.97 | -64 | 51 | 18.3 | 17.50 | 0.11 | 20.89 | 0.36 | 24.00 | — |
| 206 | 20 | 52 | 53.85 | -64 | 49 | 23.8 | 17.54 | 0.03 | 18.84 | 0.02 | 19.88 | 0.01 |
| 158 | 20 | 52 | 50.12 | -64 | 50 | 08.2 | 17.55 | 0.03 | 19.83 | 0.03 | 22.25 | 0.05 |
| 109 | 20 | 52 | 49.17 | -64 | 50 | 51.0 | 17.56 | 0.02 | 19.53 | 0.02 | 21.66 | 0.03 |
| 85 | 20 | 52 | 58.29 | -64 | 51 | 14.2 | 17.56 | 0.02 | 19.78 | 0.02 | 21.89 | 0.05 |
| *217 | 20 | 52 | 58.27 | -64 | 49 | 18.6 | 17.63 | 0.03 | 19.39 | 0.02 | 23.73 | 0.19 |
| 153 | 20 | 53 | 04.53 | -64 | 50 | 10.7 | 17.66 | 0.03 | 19.79 | 0.03 | 24.00 | — |
| 98 | 20 | 52 | 47.73 | -64 | 51 | 02.3 | 17.66 | 0.03 | 19.17 | 0.02 | 24.25 | 0.29 |
| 141 | 20 | 52 | 38.33 | -64 | 50 | 18.7 | 17.69 | 0.03 | 20.13 | 0.03 | 23.95 | 0.20 |
| 78 | 20 | 53 | 02.56 | -64 | 51 | 22.8 | 17.70 | 0.02 | 21.04 | 0.07 | 24.00 | — |
| 71 | 20 | 52 | 54.03 | -64 | 51 | 25.3 | 17.72 | 0.03 | 20.57 | 0.05 | 24.00 | — |
| 14 | 20 | 52 | 55.32 | -64 | 52 | 12.9 | 17.75 | 0.03 | 20.18 | 0.03 | 23.91 | 0.19 |
| 73 | 20 | 52 | 58.64 | -64 | 51 | 24.1 | 17.82 | 0.03 | 20.02 | 0.03 | 23.16 | 0.11 |
| 204 | 20 | 52 | 45.33 | -64 | 49 | 27.8 | 17.85 | 0.03 | 21.06 | 0.06 | 23.85 | 0.19 |
| 203 | 20 | 52 | 45.84 | -64 | 49 | 28.1 | 17.86 | 0.03 | 21.27 | 0.09 | 24.00 | — |
| 42 | 20 | 52 | 51.36 | -64 | 51 | 48.8 | 17.90 | 0.03 | 20.40 | 0.04 | 23.25 | 0.11 |
| 116 | 20 | 52 | 43.58 | -64 | 50 | 47.9 | 17.93 | 0.03 | 20.99 | 0.08 | 24.00 | — |
| 27 | 20 | 52 | 46.57 | -64 | 52 | 01.3 | 17.94 | 0.03 | 21.00 | 0.07 | 24.00 | — |
| 115 | 20 | 52 | 44.35 | -64 | 50 | 50.0 | 17.94 | 0.03 | 20.26 | 0.04 | 24.00 | — |
| 171 | 20 | 53 | 06.87 | -64 | 49 | 56.7 | 17.97 | 0.04 | 21.22 | 0.09 | 23.23 | 0.07 |
| *140 | 20 | 52 | 40.15 | -64 | 50 | 20.6 | 17.99 | 0.03 | 19.54 | 0.02 | 24.01 | 0.22 |
| *94 | 20 | 52 | 50.88 | -64 | 51 | 05.8 | 17.99 | 0.03 | 19.66 | 0.02 | 24.00 | — |

**Figure 5.** (a) Observed distribution in the $EW(H\delta)$ versus $B - R$ plane for cluster members with $R < 20$ for the 3 $z = 0.31$ clusters: AC114 (triangles), AC103 (circles), and AC118 (squares). Numbers indicate the sample sizes in the various star-formation categories. (b) Simulated distribution reproducing the data from (a) according to a mixture of 60 per cent early-type galaxies and 40 per cent spirals, of which 30 per cent are bursting according to the precepts of Fig. 4 (clear symbols: squares=spirals, triangles=spheroidals) and the remainder are non-bursters (filled symbols: squares=spirals, triangles=spheroidals). Numbers indicate the sample sizes in the various star-formation categories.

**Figure 6.** Deep $K'$−band images of AC114, AC118, and AC103 obtained with the IRIS infrared imaging camera on the Anglo-Australian Telescope. The HST fields are drawn on the images where observations are available. The numbers on the axes are in pixels; the images are $3.2'$ on a side with a $0.395''$ pixel scale. North is up and east is to the left.

**Figure 7.** Absolute magnitude distribution for the combined sample of cluster galaxies after field correction. Error bars are based on Poisson statistics. The smooth line shows the best-fit Schechter luminosity function with $M^*_{K'} = -25.23^{+0.13}_{-0.14}$ and $\alpha = -1.0^{+0.13}_{-0.11}$.

**Figure 9.** Data versus model comparison for the $U - I$ versus $I - K'$ colour-colour plane. (a) Combined data across all 3 clusters limited at $K' = 18$. Filled symbols refer to spectroscopic members. The numbers of objects expected in both the cluster and the field are given for each region. (b) Simulated distribution for the same model as in Fig. 5(b).

**Figure 8.** Observed $I - K'$ versus $K'$ and $U - I$ versus $K'$ C-M diagrams for $K' \leq 18$ mag cluster and field galaxies in the IRIS frames of (left) AC114, (middle) AC118, and (right) AC103 with the predicted no-evolution C-M relations based on the Coma cluster viewed at $z = 0.31$.

the spectroscopic data is a much finer probe of recent activity than stellar debris associated with the encounter.

It is instructive to attempt to compare the frequencies and timescales of the various dynamical stages with those inferred from the HST morphologies. AC114 morphologies based on WFPC-1 data have already been tabulated by Couch et al. (1994). As discussed above, we now have the AC118 and AC103 images from which a classification list has been derived for all three clusters in the scheme (i)-(iv) above. We chose to do this for the larger $K' \leq 18$ sample (190 objects) rather than for the spectroscopic set.

Several authors (AJB, WJC, RSE) classified the galaxies to $K' \leq 18$ in AC103, AC114 and AC118 according to the Mihos scheme. Although there were some discrepancies between authors in separating the galaxies into the active categories (i)-(iii), classifications relative to category (iv) were stable. Objects classified as either regular or disturbed spirals do not fall within the Mihos scheme and therefore could not be included in the following analysis. For both clusters combined, the number distribution in the four Mihos categories was found to be N(i) 15, N(ii) 15, N(iii) 13, and N(iv) 95. This analysis is complicated by the fact that field contamination is more likely to affect categories N(i)-N(iii) than N(iv). If mergers are widespread in faint field samples, as appears to be the case (c.f. Griffiths et al. 1994), a spurious result may arise. However, we can still address the problem qualitatively. If we assume that *all* bursting galaxies are driven by mergers whose remnants are spheroidals, then according to our model, we should expect the ratio N(i)-N(iii)/N(iv)$\simeq$ 0.5. Thus, the ratio N(i)-N(iii)/N(iv)$\simeq$ 0.44 seen in the HST images is consistent.

Two possibilities emerge. Firstly, the number of interacting galaxies may be consistent with the idea of widespread merging as the basic origin of the abnormal star-forming galaxies. The small timescales available for recognising the tell-tale morphological signatures would appear to make the detection of so many unusual sources in the 3 clusters a very significant result, notwithstanding field contamination.

Alternatively, one might interpret the numbers of SB, PSG, or morphologically peculiar galaxies as being inconsistent with a large fraction ($\simeq$ 30 per cent) of active galaxies if the driving dynamical process were *longer-lived*. Several other mechanisms have been proposed for the demise of spiral galaxies in intermediate redshift clusters, including tidal friction (Valluri 1993; Valluri, in preparation; Moore et al., in preparation). Promising though these explanations appear to be, a major difficulty will be reconciling the number of red post-burst galaxies. As this phase is measured in Gyr by the lifetime of A-type stars on the main sequence in the burst phase, the stellar clock provides a very important constraint on the number of galaxies undergoing recent activity.

## 6   CONCLUSIONS

● We have made a detailed comparison of the distributions of galaxies in objective samples selected from 3 $z = 0.31$ clusters against evolutionary models based on the concept of a cluster population, some fraction of which are in the various stages of a secondary burst of star formation. We find good agreement between the model predictions and the observed distributions in both the $H\delta$ versus $B - R$ and $U - I$ versus $I - K'$ planes. The most robust conclusion from the modelling, as constrained by the high proportion of $H\delta$-strong red galaxies, is that $\simeq$ 30 per cent of all cluster galaxies must be involved in this activity within the last $\sim$ 2Gyr prior to the epoch of observation. With less security, the model fits suggest the bursts are typically short-lived ($\simeq$0.1 Gyr) and produce $10 - 20$ per cent of a galaxy's total stellar mass.

● We examine the possibility that mergers alone drive the trends seen in these clusters. We find that the percentage of galaxies with $K' \leq 18$ mag in the combined WFPC fields of AC114 and AC118 which fall into one of the two Mihos categories comprising either galaxies exhibiting tidal tails prior to a merging event or galaxies in the process of merging is very high, notwithstanding field contamination. As these are very short-lived phases ($< 1$ Gyr), a high frequency of activity seems inevitable unless the timescales can be lengthened. In that case, a major constraint will be the number of red HDS galaxies whose evolutionary timescale is well-constrained by stellar evolution.

## ACKNOWLEDGEMENTS

We acknowledge useful discussions and encouragement from Alan Dressler and Gus Oemler. AJB acknowledges the financial support of the Marshall Aid Commemoration Commission, AAS acknowledges that of the Royal Society, and IRS acknowledges that of NATO/OCIW. WJC acknowledges the financial support of the Australian Research Council, the Australian Department of Science and Technology, and Sun Microsystems.

## REFERENCES

Allen D. 1992, *IRIS users' guide*, Anglo-Australian Observatory User Manual, 30a
Alonso-Herrero A., Aragón-Salamanca A., Zamorano J., Rego M., MNRAS, in press
Aragón-Salamanca A., Ellis R. S., Sharples R. M., 1991, MNRAS, 248, 128 (AES)
Aragón-Salamanca A., Ellis R. S., Couch W. J., Carter D., 1993, MNRAS, 262, 764 (AECC)
Beers T. C., Flynn K., Gebhardt K., 1990, AJ, 100, 32
Bower R. G., Lucey J. R., Ellis R. S., 1992, MNRAS, 254, 601
Bruzual G., 1981, PhD thesis, University of California, Berkeley
Bruzual G., 1983, ApJ, 273, 105
Bruzual G., Charlot S., 1993, ApJ, 405, 538 (BC)
Butcher H., Oemler A., 1978, ApJ, 219, 18
Butcher H., Oemler A., 1984, ApJ, 285, 426
Charlot S., Ferrari F., Mathews G. J., Silk J., 1993, ApJ, 419, L57
Cole S., Aragón-Salamanca A., Frenk C. S., Navarro J. F., Zepf S. E., 1994, MNRAS, 271, 781
Couch W. J., Newell E. B., 1984, ApJS, 56, 143
Couch W. J., Newell E. B., 1990, PASP, 92, 610
Couch W. J., Sharples R. M., 1987, MNRAS, 229, 423 (CS)
Couch W. J., Ellis R. S., Godwin J., Carter D., 1983, MNRAS, 205, 1287
Couch W. J., Ellis R. S., Sharples R. M., Smail I., 1994, ApJ, 430, 121
Cowie L. L., Gardner J. P., Hu E. M., Songaila A., Hodapp K. W., Wainscoat R. J., 1994, ApJ, 434, 114
Dressler A., 1980, ApJS, 42, 565
Dressler A., Gunn J. E., 1982, ApJ, 263, 533
Dressler A., Gunn J. E., 1983, ApJ, 270, 7
Dressler A., Oemler A., Butcher H., Gunn J. E., 1994, ApJ, 430, 107
Ellis R. S., Couch W. J., MacLaren I., Koo D. C., 1985, MNRAS, 217, 239
Gardner J. P., Cowie L. L., Wainscoat R. J., 1994, ApJ, 415, L9

Glazebrook K., Peacock J. A., Miller L., Collins C. A., 1995, MNRAS, in press

Graham J. A., 1982, PASP, 94, 244

Griffiths R. E. et al., 1994, ApJ, 435, L19

Irwin M. J., 1985, MNRAS, 214, 575

Johnson H. L., 1966, ARA&A, 4, 193

Koornneef J., 1983, A&A, 128, 84

Lavery R. J., Henry J. P., 1986, ApJ, 304, L5

Loveday J., Peterson B. A., Efstathiou G., Maddox S. J., 1992, ApJ, 390, 338

MacLaren I., Ellis R. S., Couch W. J., 1988, MNRAS, 230, 249

Mihos J. C., 1995, ApJ, 438, L75

Mobasher B., Sharples R. M., Ellis R. S., 1993, MNRAS, 263, 560

Schechter P. L., 1976, ApJ, 203, 297

Valluri M., 1993, ApJ, 408, 57

Wirth G. D., Koo D. C., Kron R. G., 1994, ApJ, 435, L105